\documentclass[preprint,prd,superscriptaddress,amsmath,nofootinbib]{revtex4}

\pdfoutput=1

\usepackage{amsmath}
\usepackage{graphicx}
\usepackage{dcolumn}
\usepackage[hyperfootnotes=false]{hyperref}
\usepackage{xspace}
\usepackage{etoolbox}
\usepackage{enumerate}
\usepackage{enumitem}
\usepackage[caption=false]{subfig}

\renewcommand{\Re}{\text{Re}\,}

\newcommand{\ga}{g_{\rm A}}
\newcommand{\Order}{O} 
\newcommand{\MeV}{\,\text{MeV}}

\newcommand{\sone}{Method~1}
\newcommand{\stwo}{Method~2}
\newcommand{\sthree}{Method~3}

\allowdisplaybreaks[1]

\begin{document}
\preprint{UWThPh-2015-5}
\preprint{CERN-PH-TH-2015-047}

\title{Light stops, blind spots, and isospin violation in the MSSM}
\author{Andreas Crivellin}
\affiliation{CERN Theory Division, CH--1211 Geneva 23, Switzerland}
\affiliation{Albert Einstein Center for Fundamental Physics,
Institute for Theoretical Physics,\\ University of Bern, Sidlerstrasse 5, CH--3012 Bern, Switzerland}
\author{Martin Hoferichter}
\affiliation{Institut f\"ur Kernphysik, Technische Universit\"at Darmstadt, D--64289 Darmstadt, Germany}
\affiliation{ExtreMe Matter Institute EMMI, GSI Helmholtzzentrum f\"ur Schwerionenforschung GmbH, D--64291 Darmstadt, Germany}
\affiliation{Albert Einstein Center for Fundamental Physics,
Institute for Theoretical Physics,\\ University of Bern, Sidlerstrasse 5, CH--3012 Bern, Switzerland}
\author{Massimiliano Procura}
\affiliation{Fakult\"at f\"ur Physik, Universit\"at Wien, Boltzmanngasse 5, A--1090 Vienna, Austria}
\affiliation{Albert Einstein Center for Fundamental Physics,
Institute for Theoretical Physics,\\ University of Bern, Sidlerstrasse 5, CH--3012 Bern, Switzerland}
\author{Lewis C.\ Tunstall}
\affiliation{Albert Einstein Center for Fundamental Physics,
Institute for Theoretical Physics,\\ University of Bern, Sidlerstrasse 5, CH--3012 Bern, Switzerland}

\makeatletter
\patchcmd{\frontmatter@RRAP@format}{(}{}{}{}
\patchcmd{\frontmatter@RRAP@format}{)}{}{}{}
\renewcommand\Dated@name{}
\makeatother


\begin{abstract}
In the framework of the MSSM, we examine several simplified models where only a 
few superpartners are light. This allows us to study WIMP--nucleus scattering in 
terms of a handful of MSSM parameters and thereby scrutinize their impact on dark 
matter direct-detection experiments. Focusing on spin-independent WIMP--nucleon 
scattering, we derive simplified, analytic expressions for the Wilson coefficients 
associated with Higgs and squark exchange. We utilize these results to study the 
complementarity of constraints due to direct-detection, flavor, and collider 
experiments. We also identify parameter configurations that produce (almost) 
vanishing cross sections. In the proximity of these so-called blind spots, 
we find that the 
amount of isospin violation may be much larger than typically expected 
in the MSSM. This feature is a generic property of parameter regions where cross 
sections are suppressed, and highlights the importance of a careful analysis of 
the nucleon matrix elements and the associated hadronic uncertainties.  This becomes 
especially relevant once the increased sensitivity of future direct-detection 
experiments corners the MSSM into these regions of parameter space.
\end{abstract}

\maketitle

\section{Introduction}
\label{intro}
Establishing the microscopic nature of Dark Matter (DM) is one of the central, 
open questions in cosmology and particle physics. In the context of cold nonbaryonic 
DM, the prevailing paradigm is based on weakly interacting massive particles (WIMPs), 
and extensive theoretical and experimental resources have been devoted towards 
identifying viable candidates and developing methods to detect them.  One of the 
most studied WIMPs arises in the Minimal Supersymmetric Standard Model (MSSM), 
where an assumed $R$-parity ensures that the lightest superpartner (LSP) is a 
stable neutralino $\chi$ composed of bino $\tilde{B}$, wino $\tilde{W}$, and 
Higgsino $\tilde{H}$ eigenstates. The mass of the LSP is expected to lie in the 
range of tens to hundreds of GeV.

In its general form, however, the MSSM contains more than $100$ parameters, most 
of which are tied to the hidden sector which breaks supersymmetry (SUSY) at some 
scale $M_\mathrm{SUSY}$.  Since these parameters are unknown {\it a priori}, it 
is necessary to restrict the dimensionality of the parameter space in order to 
obtain a predictive framework with which to undertake phenomenological analyses. 
One way to achieve this is to adopt a specific mechanism that describes high-scale 
SUSY-breaking in terms of a small number of parameters. For example, the constrained 
MSSM (CMSSM) with minimal supergravity~\cite{Chamseddine:1982jx,Nath:1982zq,Barbieri:1982eh,Hall:1983iz} only involves five free parameters, but faces increasing tension~\cite{Buchmueller:2011ab,Buchmueller:2012hv,Cabrera:2012vu,Strege:2012bt,Fowlie:2012im,Buchmueller:2013rsa,Bechtle:2012zk,Bechtle:2013mda} with the non-observation of superpartners at the 
LHC experiments and other observables like the measured Higgs mass and anomalous 
magnetic moment of the muon. Alternatively, one can remain agnostic about the 
features of SUSY-breaking and incorporate data-driven constraints, as in e.g.\ 
the p(henomenological)MSSM~\cite{Djouadi:1998di,Berger:2008cq}, where only 19 free 
parameters are used to capture the essential features of weak-scale SUSY.

In both approaches, long computational chains involving spectrum generators, the 
calculation of decay rates, or the DM relic abundance are typically required in 
order to explore the relevant parameter space.  This strategy has been used 
extensively for the CMSSM~\cite{Buchmueller:2011ab,Buchmueller:2012hv,Cabrera:2012vu,Strege:2012bt,Fowlie:2012im,Buchmueller:2013rsa,Bechtle:2012zk,Bechtle:2013mda} and pMSSM~\cite{Boehm:2013qva,Grothaus:2012js,Han:2013gba,Strege:2014ija} to 
analyze $\chi$--nucleon scattering and impose limits from current DM direct-detection 
experiments such as SuperCDMS~\cite{Agnese:2014aze}, XENON100~\cite{Aprile:2011hi}, 
and LUX \cite{Akerib:2013tjd}, as well as upcoming proposals like XENON1T~\cite{Aprile:2012zx}, 
LUX-ZEPLIN (LZ)~\cite{Malling:2011va}, and SuperCDMS SNOLAB~\cite{Brink:2012zza}. 
While these parameter scans allow one to gain useful information about the status 
of the theory in light of global fits, they generally hinder attempts to clearly 
identify which contributions associated with the underlying theory parameters can 
have the greatest impact on a signal of interest.  An {\it analytical} understanding 
of the underlying parameter space can instead be obtained in the context of so-called 
simplified models, defined%
	\footnote{For a definition of ``simplified models'' in the context of LHC searches, 
	see~\cite{Alwall:2008ag,Alves:2011wf,Barnard:2014joa,Edelhauser:2014ena,Calibbi:2014lga}.}
\cite{Cheung:2012qy} to be minimal theories of weak-scale SUSY where all but 
a handful of the superpartners relevant for DM phenomenology are decoupled from 
the spectrum.

For spin-independent (SI) $\chi$--nucleon scattering, the choice of simplified 
model is guided by the dominant contributions to the cross section, namely, Higgs 
and squark exchange~\cite{Goodman:1984dc,Griest:1988ma,Srednicki:1989kj,Giudice:1988vs,Shifman,Drees:1992rr}.  
To date, the focus has largely concerned the role of the Higgs sector, both in 
the decoupling limit where a single SM-like Higgs $h$ is present in the 
spectrum~\cite{Cheung:2012qy,Hisano:2012wm}, or in the more general 
case~\cite{Huang:2014xua,Anandakrishnan:2014fia} where the heavier $CP$-even Higgs 
$H$ is included.  This focus is chiefly motivated by the fact that current bounds 
on the masses of gluinos and (degenerate) squarks of the first two generations are 
larger than about 1 TeV~\cite{Aad:2014wea,Chatrchyan:2014lfa}, and so their 
contribution to the SI cross section can be safely ignored.%
	\footnote{However, for non-degenerate squarks~\cite{Nir:1993mx,Nir:2002ah} 
	the constraints from FCNCs are satisfied~\cite{Crivellin:2010ys}, and the 
	collider bounds are significantly weakened~\cite{Mahbubani:2012qq}.  In 
	this case, contributions from the first two generations could also be important 
	for SI $\chi$--nucleon scattering.}

However, the decoupling of third-generation squarks---especially stops---upsets 
the main motivation behind the MSSM, namely, its ability to stabilize the 
electroweak scale $v \simeq$~174~GeV against loop corrections in a technically 
natural fashion~\cite{Dimopoulos:1995mi,Giudice:2004tc,Kitano:2006gv,Perelstein:2007nx,Brust:2011tb,Papucci:2011wy}.  
In other words, if naturalness is to remain a useful criterion with which to 
constrain the MSSM parameter space, then the spectrum should (minimally) include 
light stops $\tilde{t}_{1,2}$ and---due to $SU(2)_L$ invariance---a left-handed  
sbottom $\tilde{b}_L$~\cite{Barbieri:1987fn,Dimopoulos:1995mi,Cohen:1996vb,Barbieri:2009ev}. 
While the search for top and bottom squarks remains a primary focus of the LHC 
experiments, their impact on DM direct-detection limits has not been explored in 
detail.

The purpose of this paper is to present an analytical scheme which allows one to 
successively include those states which are most relevant for naturalness and DM 
direct detection. In particular, we consider a bino-like LSP and derive 
simplified, analytic expressions for the Wilson coefficients associated with SI 
scattering.  We examine in detail the contributions from Higgs and third-generation 
squark exchange, and study the interplay of collider, flavor, and DM constraints. 
As in previous analyses of the Higgs 
sector~\cite{Cheung:2012qy,Hisano:2012wm,Huang:2014xua,Anandakrishnan:2014fia}, 
our scheme allows us to identify so-called blind spots in parameter space, where 
the SI cross section is strongly suppressed by either a particular set of 
parameters~\cite{Hisano:2012wm,Cheung:2012qy}, or destructive interference~\cite{Huang:2014xua,Anandakrishnan:2014fia} in the scattering amplitude.  This effect was first identified 
numerically through a scan of the CMSSM parameter space~\cite{Ellis:2000ds,Ellis:2000jd,Baer:2006te}, while lower bounds on the $\chi$--nucleon cross sections were first discussed 
in~\cite{Mandic:2000jz} for both the CMSSM and a generalized MSSM framework. 
A key feature of our analysis is that in the vicinity of blind spots, the 
amount of isospin violation may be much larger than typically expected for the 
MSSM~\cite{Ellis}.%
	\footnote{Large isospin violation has been put forward as a mechanism to 
	reconcile contradictory DM direct-detection signal claims and null observations 
	\cite{Kurylov:2003ra,Giuliani:2005my,Chang:2010yk,Feng:2011vu,Cirigliano:2012pq,Cirigliano13}.} 
In order to account for isospin-violating effects originating from the
 nucleon matrix elements of the scalar quark currents, we use the formalism developed in~\cite{Crivellin:2013ipa}, which 
provides an accurate determination of the hadronic uncertainties.

The paper is organized as follows. In Sec.~\ref{sec:theory} we establish our 
notation for $\chi$--nucleus scattering and comment on the differing treatments 
\cite{Ellis,Belanger:2013oya,Crivellin:2013ipa} of the nucleon scalar matrix elements 
found in the literature. The leading MSSM contributions to the Wilson coefficients 
are then identified using a systematic expansion in $v/M_{\rm SUSY}$ 
which generates simplified analytic expressions for the Wilson coefficients 
associated with Higgs and squark exchange.  Section~\ref{sec:simplified models} 
examines four simplified models, where, driven by naturalness, we successively 
include the most relevant particles as active degrees of freedom. 
In each case, we discuss the conditions for blind spots and examine the amount of 
isospin violation allowed by current and projected limits from SI DM scattering. 
Our analysis shows that the absence of DM signals pushes the MSSM into regions of 
parameter space where isospin-violating effects are likely to become relevant. 

\section{Theoretical Preliminaries}
\label{sec:theory}

\subsection{Spin-independent neutralino--nucleus cross section: scalar matrix elements}
\label{sec:hadronic}
We start by providing some definitions for the elastic scattering of the lightest 
neutralino $\chi$ off a species of nucleus ${\cal N}= {}^A_ZX$, where $Z$ and $A$ 
denote the atomic and mass numbers respectively. Typically, the dependence of the 
cross section on the small momentum transfer is assumed to be described by nuclear 
form factors.  At zero momentum transfer and for one-body currents only, the 
cross section for $\chi {\cal N} \to \chi {\cal N}$ is given by
\begin{equation}
\label{eq:cross}
\sigma_\mathrm{SI} = \frac{4\mu_\chi^2}{\pi} [Zf_p + (A-Z)f_n]^2\,.
\end{equation}
Here, $\mu_\chi = m_\chi m_{\cal N}/(m_\chi + m_{\cal N})$ is the reduced mass of 
the $\chi$--${\cal N}$ system, while $f_p$ and $f_n$ are effective (zero-momentum) 
SI couplings of the LSP to the proton and neutron respectively.

For nucleons $N$, the $\chi$--$N$ couplings $f_N$ are defined by
\begin{equation}
\label{fN}
\frac{f_N}{m_N} = \sum_{q=u,d,s} f_{q}^N C_{q} 
+ \,f_{Q}^N\sum_{q=c,b,t} C_{q}\,,
\qquad N=p \mbox{ or } n\,, 
\end{equation}
where $C_{q}$ is the Wilson coefficient of the scalar operator 
$\bar{m}_{q}\bar{\chi}\chi\,\bar{q}q$ with running quark mass $\bar{m}_q$, and 
\begin{equation}
\label{fsum}
	m_N f^N_q  = \langle N|\bar m_q \bar{q}q|N\rangle\,, 
	\qquad f_{Q}^N = \frac{2}{27}(1-f_u^N - f_d^N - f_s^N)\,.
\end{equation}
The coefficients $f^N_q$ can be interpreted as the fraction of the nucleon mass generated 
by the respective quark scalar current and are often referred to as nucleon scalar couplings. 
In the framework adopted in~\eqref{fsum}, the heavy quarks $c,b,t$ are integrated out, so that, via the 
trace anomaly~\cite{Adler:1976zt,Minkowski:1976en,Nielsen:1977sy,Collins:1976yq} 
of the QCD energy-momentum tensor, their scalar coefficients $f_{Q}^N$ can be expressed in 
terms of the light-quark ones~\cite{Shifman}. As shown by Drees and 
Nojiri~\cite{Drees:1993bu}, this procedure fails if the squarks are sufficiently 
light, and in Sec.\ \ref{sec:theory} we discuss the necessary modifications to 
(\ref{fN}) which account for the exact one-loop result. 

We note that~\eqref{fsum} holds at leading order in $\alpha_s$: in the case of 
the charm quark, this may not be sufficiently accurate, so that either higher-order 
corrections~\cite{Kryjevski,Vecchi,Hill:2014yxa} or a non-perturbative determination on the 
lattice could become mandatory. Similarly, corrections to the single-nucleon picture 
underlying~\eqref{eq:cross} in the form of two-nucleon currents can be systematically 
taken into account using effective field theory~\cite{Prezeau:2003sv,Cirigliano:2012pq,Cirigliano13,Hoferichter:2015ipa}. In this paper, we use (\ref{eq:cross}) and (\ref{fsum}) to investigate the amount 
of isospin violation that can be generated within several simplified models, given 
the hadronic uncertainties of the single-nucleon coefficients $f^N_q$ for the light 
quarks $u,d,s$. 

Traditionally, the scalar matrix elements of the light quarks have been 
determined from a combination of chiral $SU(3)_L\times SU(3)_R$ perturbation theory 
($\chi$PT$_3$) and phenomenological input inferred from the pion--nucleon $\sigma$-term 
$\sigma_{\pi N}$ and the hadron mass spectrum~\cite{Ellis:2000ds,Corsetti:2000yq,Ellis,Cirigliano:2012pq}.
A central feature of this approach is that the up- and down-quark coefficients
$f_{u,d}^N$ are reconstructed from two three-flavor quantities: the so-called 
strangeness content of the nucleon  
\begin{equation}
	y = \frac{2\langle N|\bar{s}s|N\rangle}{\langle N|\bar{u}u+\bar{d}d|N\rangle}
	= \frac{2f_s^N/\bar{m}_s}{f_u^N/\bar{m}_u + f_d^N/\bar{m}_d}\,,
\end{equation}
and another parameter  
\begin{equation}
	z = \frac{\langle N| \bar{u}u-\bar{s}s|N\rangle}{\langle N|\bar{d}d-\bar{s}s|N\rangle}
	= \frac{f_u^N/\bar{m}_u - f_s^N/\bar{m}_s}{f_d^N/\bar{m}_d - f_s^N/\bar{m}_s}
\end{equation}
that is related 
to isospin violation.  As a result, the inherent uncertainties of $\chi$PT$_3$ 
(typically of order $30\%$) propagate to the two-flavor sector. Furthermore, $z$ is 
usually extracted from a leading-order fit to baryon masses~\cite{Cheng:1988im}, 
and this compounds the problem of obtaining reliable uncertainty estimates. 

For the up- and down-quark coefficients $f_{u,d}^N$, these problems can be circumvented 
by using the two-flavor theory $\chi$PT$_2$ directly, thus avoiding the 
three-flavor expansion in the first place~\cite{Crivellin:2013ipa}. Starting from 
the $\chi$PT$_2$ expansion of the nucleon mass at third chiral order and including
the effects due to strong isospin violation, one finds
\begin{align}
\label{fufd}
	 f_u^N &= \frac{\sigma_{\pi N}(1-\xi)}{2m_N} + \Delta f_u^N\,, & f_d^N &= \frac{\sigma_{\pi N}(1+\xi)}{2m_N} + \Delta f_d^N\,, \notag \\
	 \Delta f_u^p &= (1.0\pm 0.2) \times 10^{-3}\,,  &\Delta f_u^n &= (-1.0\pm0.2)\times 10^{-3}\,,\notag \\
	 \Delta f_d^p &= (-2.1\pm 0.4) \times 10^{-3}\,, & \Delta f_d^n &= (2.0\pm0.4)\times 10^{-3}\,,
\end{align}
where the $\sigma$-term is defined as $\sigma_{\pi N}\equiv \langle N|\hat m(\bar{u}u+\bar{d}d)|N\rangle$, 
averaged over proton and neutron, 
$\hat m=(\bar m_u+\bar m_d)/2$, and
\begin{equation}
\xi = \frac{\bar m_d-\bar m_u}{\bar m_d+\bar m_u} = 0.36 \pm 0.04
\end{equation}
is taken from~\cite{FLAG}.

For the present work, one particularly important aspect of the $\chi$PT$_2$ 
approach~\cite{Crivellin:2013ipa} is that isospin violation can be rigorously 
accounted for, including uncertainty estimates. This aspect can be nicely 
illustrated by considering the differences
\begin{equation}
f_u^p - f_u^n = (1.9\pm 0.4)\times 10^{-3} \qquad \mbox{and} \qquad 
f_d^p-f_d^n = (-4.1\pm0.7)\times 10^{-3}\,,
\end{equation}
wherein the terms $\sigma_{\pi N}(1\pm\xi)/2m_N$ from~\eqref{fufd} cancel.%
	\footnote{The chiral expansion of the nucleon mass difference $m_p-m_n$ is 
	known to have a large chiral logarithm at fourth order, with coefficient 
	$(6\ga^2+1)/2\approx 5$~\cite{Frink:2004ic,Tiburzi:2005na}. We have checked that 
	including this logarithm in the analysis leads to changes of the 
	$\Delta f_{u,d}^N$ well within the uncertainties given in~\eqref{fufd}.} 
Using the $\chi$PT$_3$ approach, these differences are overestimated by roughly a factor 
of $2$, as in e.g.~\cite{Belanger:2013oya}:
\begin{equation}
\label{IV_Micromegas}
 f_u^p-f_u^n=4.3\times 10^{-3}\,,\qquad
 f_d^p-f_d^n=-8.2\times 10^{-3}\,.
\end{equation}
Alternatively, one could introduce further measures of isospin violation like  
$f_u^p-f_d^n$ and $f_u^n-f_d^p$ (motivated by the quark-model picture of the 
nucleon), but these combinations depend on the specific value of $\sigma_{\pi N}$. 
In the isospin-conserving limit, all up- and down-coefficients obtained from the chiral expansion of 
the nucleon mass become equal $f_u^p=f_u^n=f_d^p=f_d^n=\sigma_{\pi N}/2m_N$, so 
that the relations $f_u^p=f_d^n$ and $f_u^n=f_d^p$ are fulfilled. 

Ultimately, the quantities relevant for the direct-detection cross section are 
the parameters defined in~\eqref{fN}, after multiplication by the Wilson coefficients 
and summing over quark flavors. In particular, the cross section~\eqref{eq:cross} 
may be rewritten as
\begin{equation}
\label{cross_IV}
\sigma_\mathrm{SI} = \frac{4\mu_\chi^2}{\pi}f_p^2 \bigg[A + (A-Z)\bigg(\frac{f_n}{f_p}-1\bigg)\bigg]^2\,, 
\end{equation}
so that the departure of $f_n/f_p$ from unity emerges as a convenient measure of 
isospin violation. In this context, care has to be taken in interpreting the limits 
on the WIMP--{\it nucleon} cross section $\sigma_\mathrm{SI}^{p,n}$ given by experimental collaborations. Indeed, these are generally extracted via the relation 
\begin{equation}
	\sigma_\mathrm{SI} = \sigma_\mathrm{SI}^N \left(\frac{\mu_\chi}{\mu_N}\right)^2 A^2\,, 
	\label{eqn:cross SI}
\end{equation}
where $\mu_N$ is the reduced mass of the $\chi$--nucleon system. We stress that $\sigma_\mathrm{SI}^{p}$ in (\ref{eqn:cross SI}) can be identified with the SI $\chi$--proton cross section only under the {\it assumption} $f_n\simeq f_p$. If isospin-violating effects are large, it is natural to compare against the $\chi$--nucleus cross section (\ref{eq:cross}) directly, and (\ref{eqn:cross SI}) indicates how the experimental limits are to be rescaled.

\begin{figure}[t]
	\centering\includegraphics[scale=0.8]{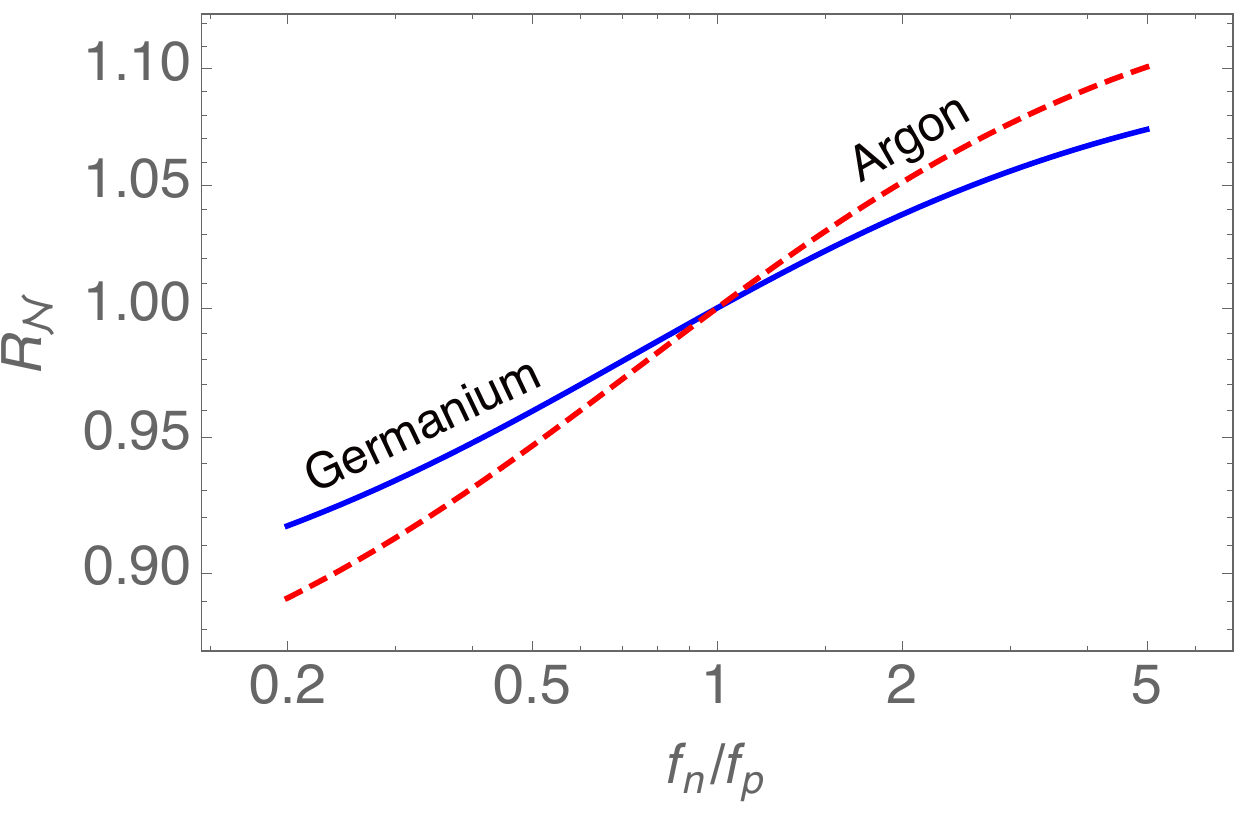}
	\caption{Relative difference $R_{\cal N}$ (defined in (\ref{xe vs nuclei})) between xenon and other nuclear targets ${\cal N}$ as a function of the isospin violation measure $f_n/f_p$.  The dashed line in red shows the comparison against ${\cal N} = \text{germanium}$, while the solid blue line corresponds to the ${\cal N}=\text{argon}$ comparison.}
	\label{fig:target IV}
\end{figure}

In general, the effects of isospin violation depend on the target nucleus. For the mass range $m_\chi \simeq 50$-$1000$ GeV considered in most of our analysis (Sec.~\ref{sec:simplified models}), the strongest limits on SI $\chi$--nucleon scattering are currently set by LUX~\cite{Akerib:2013tjd}.  In the context of isospin violation, this prompts us to focus on the projected reach of upcoming xenon-based experiments like XENON1T~\cite{Aprile:2012zx} and LZ~\cite{Malling:2011va}.  However, this raises the question whether other experiments like SuperCDMS SNOLAB~\cite{Brink:2012zza} (based on germanium) can be used to place complementary constraints on $f_n/f_p$.  To quantify the difference between xenon-based constraints and other nuclei, consider the $f_n/f_p$ dependence of the ratio 
\begin{equation}
	R^{}_{\cal N} = \frac{\sigma_\mathrm{SI}^\mathrm{Xe}}{\sigma_\mathrm{SI}^{\cal N}} \left( \frac{\mu_\chi^{\cal N} A_{\cal N}}{\mu_\chi^\mathrm{Xe}A_\mathrm{Xe}} \right)^2\,,
	\label{xe vs nuclei}
\end{equation}
normalized such that $R_{\cal N} =1$ in the isospin-conserving limit.  For SI scattering off argon and germanium, the result is shown in Fig.\ \ref{fig:target IV}, where we observe a maximum difference of around 10\% for $f_n/f_p$ much larger or smaller than unity.  In the simplified models considered in Sec.~\ref{sec:simplified models}, the difference between $f_n$ and $f_p$ is generated entirely by SM quantities, so the improved limits offered by e.g.\ SuperCDMS SNOLAB are limited to the percent level.  Moreover, the location of blind spots is determined by the condition $f_{n,p}\simeq 0$, so neither the blind spot nor the uncertainty on $f_n/f_p$ depends significantly on the atomic/mass numbers of the nuclear target.

The crucial input quantity $\sigma_{\pi N}$ is not yet precisely determined:  in 
Sec.~\ref{sec:simplified models} we show the dependence on this parameter explicitly 
in the case of generic Higgs exchange, and later fix its central value to 
$\sigma_{\pi N}=50\MeV$ for illustrative purposes. The need for a precise determination 
of $\sigma_{\pi N}$ has triggered many ongoing efforts, including lattice-QCD 
calculations at (nearly) physical values of the pion mass; see~\cite{Young,Kronfeld:2012uk,WalkerLoud,Belanger:2013oya} for a compilation of recent results and improved 
phenomenological analyses. The challenge in the phenomenological approach, i.e.\ 
extracting $\sigma_{\pi N}$ from $\pi N$ scattering, lies in controlling the 
required analytic continuation of the isoscalar $\pi N$ amplitude  into the 
unphysical region~\cite{ChengDashen}, which might even be sensitive to 
isospin-breaking corrections~\cite{HKM,HKMlong}. This analytic continuation can 
be stabilized with the help of the low-energy data that have become available in 
recent years thanks to accurate pionic-atom measurements~\cite{Gotta:2008zza,Strauch:2010vu}, leading to a precise extraction of the $\pi N$ scattering lengths~\cite{piD,piDlong}. 
A systematic analysis of $\pi N$ scattering based on this input as well as 
constraints from unitarity, analyticity, and crossing symmetry along the lines of~\cite{RS,RSSFF,HKMR} will help clarify 
the situation concerning the phenomenological determination of $\sigma_{\pi N}$ 
\cite{Gasser:1990ce,Pavan:2001wz,Alarcon:2011zs}.

In our numerical analysis, we compare three different methods used to determine 
the scalar couplings $f_q^N$ and their uncertainties:
\begin{itemize}
\item[--] \underline{\textbf{\sone:}}  Based on $\chi$PT$_2$~\cite{Crivellin:2013ipa}, 
with $f_{u,d}^{N}$ determined from (\ref{fufd}) and $f_s^N$ from lattice QCD.  
It is well known~\cite{Giedt:2009mr} that the $\chi$--nucleon cross section is 
sensitive to the value of $f_s^N$. In our analysis we adopt the lattice average 
from~\cite{WalkerLoud}:
\begin{equation}
\label{fs_lattice}
f_s^N=0.043\pm 0.011\,.
\end{equation}
\item[--] \underline{\textbf{\stwo:}}  Corresponds to the traditional $\chi$PT$_3$ 
approach~\cite{Ellis:2000ds,Corsetti:2000yq,Ellis,Cirigliano:2012pq}, where 
$f_{u,d}^{N}$ and $f_s^N$ are determined via the three-flavor quantities $y$ and 
$z$.  In this approach, the strange-quark scalar matrix element is defined via
\begin{equation}
\label{yfs}
f_s^N= \frac{\sigma_{\pi N}}{2m_N} \frac{\bar m_s}{\hat m}y\,,
\end{equation}
where $\bar m_s/\hat m=(27.4\pm 0.4)$~\cite{FLAG}, the strangeness content is 
taken from the relation $y=1-\sigma_0/\sigma_{\pi N}$, with $\sigma_0=(36\pm 7)\MeV$ 
\cite{Borasoy}, and $z \simeq 1.49$ is extracted from leading-order fits to the baryon mass spectrum~\cite{Cheng:1988im}.
This approach introduces uncertainties that are difficult to 
quantify and is sensitive to the precise value of $\sigma_{\pi N}$. 
The range $\sigma_{\pi N}=(50\pm 15)\MeV$ covering the determinations discussed 
above~\cite{Young,Kronfeld:2012uk,Gasser:1990ce,Pavan:2001wz}
translates to $f_s^N=0.2\pm 0.2$.  Even for moderate values of the $\sigma$-term, 
large values $f_s^N\approx 0.25$ have been inferred in this way.
Such large values are incompatible with recent lattice calculations,   
which provide a more reliable determination of $f_s^N$ (see~\eqref{fs_lattice} for a recent average).  
A determination of the 
uncertainty bands arising from this approach requires us to attach an error to 
$z$, which, as argued before, is impossible to quantify reliably. Therefore, 
based on general expectations for the convergence pattern in $\chi$PT$_3$, we 
simply attach to $z$ a $30\%$ error.%
	\footnote{This is consistent with an analysis 
	\cite{Shanahan:2013cd} of the quark-mass dependence of octet baryons.}
\item[--] \underline{\textbf{\sthree:}}  Corresponds to the implementation in 
\textsc{micrOMEGAs-4.1.2}~\cite{Belanger:2013oya} and follows the traditional 
approach in~\stwo, but with~(\ref{yfs}) inverted so that $y$ is a function of 
$f_s^N$.  With lattice QCD input (\ref{fs_lattice}) for $f_s^N$, this method has 
reduced uncertainties compared with~\stwo, but suffers from the fact that 
$f_{u,d}^N$ still depend on the three-flavor quantities $y$ and $z$.  
\end{itemize}

\subsection{Simplified expressions for spin-independent scattering of bino-like dark matter}
\label{sec:susy}
Let us now derive analytic expressions for SI $\chi$--nucleon scattering in the 
MSSM. We first review the complete expressions due to tree-level Higgs and squark 
exchange, and then simplify them by expanding in powers of $v/M_{\rm SUSY}$. For 
light third-generation squarks, a procedure~\cite{Belanger:2008sj} to extend our 
results to include the one-loop corrections~\cite{Drees:1993bu} is discussed below. 

The lightest neutralino is a linear combination of $\tilde{B}$, $\tilde{W}$, and 
$\tilde{H}_{u,d}$ interaction eigenstates,
\begin{equation}
	\chi \equiv \tilde\chi^0_1 = Z^{\chi}_{11} \tilde{B} + Z^\chi_{21}\tilde{W} 
	+ Z^\chi_{31}\tilde{H}_d + Z^\chi_{41}\tilde{H}_u\,,
	\label{chi}
\end{equation}
while the neutralino mass matrix is given by
\begin{align}		
M^{\chi}=\left(\begin{array}{cccc}			
			M_1										&	0 											&	-\tfrac{1}{2}g_1v_d		&	\tfrac{1}{2}g_1v_u		\\
			0 										&	M_2											&	\tfrac{1}{2}g_2v_d		&	-\tfrac{1}{2}g_2v_u	\\
			-\tfrac{1}{2}g_1 v_d	&	\tfrac{1}{2}g_2 v_d			&	0 										&	-\mu 			\\
			\tfrac{1}{2}g_1 v_u		&	-\tfrac{1}{2}g_2v_u			&	-\mu 									&	0 					
	\end{array}\right)	\,. \label{eqn:mX}
\end{align}
Here $M_{1}$ ($M_2$) are the soft SUSY-breaking masses of the bino (wino), $\mu$ 
is the Higgsino mass parameter, and $v_{u,d}$ are the two Higgs $H_{u,d}$ vacuum 
expectation values, whose ratio $v_u/v_d$ is denoted by $\tan\beta$. Note that 
while $v_{u,d}$ can be rendered real and positive by an appropriate phase 
redefinition of the Higgs fields, $M_1$ and $M_2$ are in general complex if the 
gluino mass is assumed to be real (as is standard convention).  The neutralino 
mixing in (\ref{chi}) is determined by the unitary matrix $Z^\chi$ 
which diagonalizes $M^\chi$~\cite{Rosiek:1995kg}:
\begin{equation}
Z^\chi_{I^\prime I} M^\chi_{I^\prime J^\prime} Z^\chi_{J^\prime J}=\delta_{IJ} m_{\tilde\chi_{I}^0}\,.
\end{equation}

In the squark sector, the squared masses $m_{\tilde{q}_{s}}^2$ are eigenvalues 
of the $6\times 6$ matrices in flavor/chirality space,
\begin{align}
	{\cal M}_{\tilde{u}}^2 &= \left(\begin{array}{cc}		
		V^\dagger{\bf m}_Q^2 V + v_u^2 {\bf Y}_u{\bf Y}_u^\dagger + g_{u_L}m_Z^2c_{2\beta} 
		& - v_u({\bf Y}_u{\bf A}_u + {\bf Y}_u\mu\cot\beta) \\
		- v_u({\bf A}_u^\dagger{\bf Y}_u^\dagger + {\bf Y}^\dagger_u \mu^*\cot\beta) 
		& {\bf m}_U^2 + v_u^2{\bf Y}_u^\dagger{\bf Y}_u + g_{u_R}m_Z^2c_{2\beta}
	\end{array}	\right) \,, \notag	 \\
	{\cal M}_{\tilde{d}}^2 &= \left(\begin{array}{cc}
		{\bf m}_Q^2 + v_d^2{\bf Y}_d{\bf Y}_d^\dagger + g_{d_L}m_Z^2c_{2\beta} 
		& - v_d({\bf Y}_d{\bf A}_d + {\bf Y}_d\mu\tan\beta) \\
		- v_d({\bf A}_d^\dagger{\bf Y}_d^\dagger + {\bf Y}_d^\dagger \mu^*\tan\beta) 
		& {\bf m}_D^2 + v_d^2{\bf Y}_d^\dagger{\bf Y}_d + g_{d_R}m_Z^2c_{2\beta}
	\end{array} \right)\,.
	\label{eq:sqrk mass}
\end{align}
Here, the soft SUSY-breaking squark masses are ${\bf m}_Q, {\bf m}_U,$ and ${\bf m}_D$, 
${\bf Y}_{u,d}$ are complex Yukawa matrices, $V$ is the CKM matrix, and we have 
assumed flavor universality ${\bf a}_q = {\bf Y}_q {\bf A}_q$ for the trilinear 
$A$-terms.  We also use $s$ and $c$ for sine and cosine, so that 
$s_\beta \equiv \sin\beta$, $c_{2\beta}\equiv \cos2\beta$, etc. The weak 
neutral-current couplings
\begin{equation}
	g_q = I_3^q - e_q s_W^2
\end{equation}
are defined in terms of the third component of weak isospin $I_3^q$, electric 
charge $e_q$, and $s_W^2 \simeq 0.2231$. A unitary transformation
\begin{equation}
	Z^{\tilde{q}*}_{s^\prime s} ({\cal M}^2_{\tilde{q}})_{s^\prime t^\prime}
	Z^{\tilde{q}}_{t^\prime t} = m_{\tilde{q}_{s}}^2\delta_{st}\,,
\end{equation}
gives the physical basis with diagonal squark mass matrices, where we adopt the 
convention to order the states in increasing mass. We have also defined the 
super-CKM basis in~\eqref{eq:sqrk mass} as the one with 
diagonal (and in general, complex) Yukawa couplings $Y_{q_i}$~\cite{Crivellin:2008mq}.

\begin{figure}[t]
	\center{\includegraphics[scale=0.75]{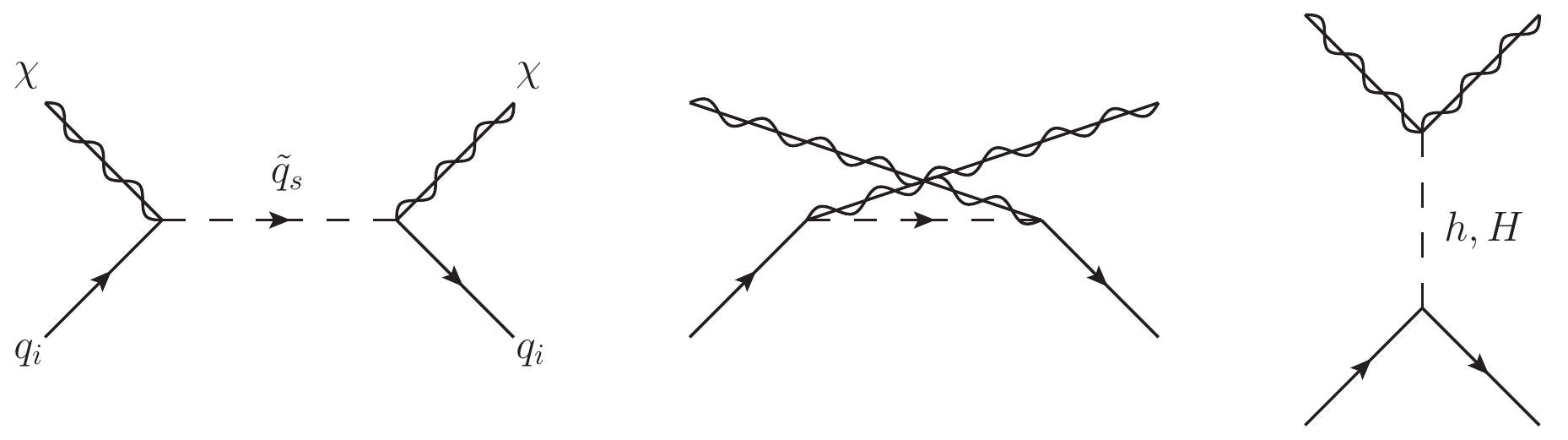}}
	\caption{Tree-level MSSM graphs which contribute to the SI  
	cross section for $\chi$--quark scattering.}
	\label{fig:spin_ind}
\end{figure}

We have now introduced the necessary ingredients to discuss $\chi$--quark scattering 
in the MSSM.  The tree-level contributions to the Wilson coefficients $C_{q_i}$ 
are the diagrams shown in Fig.~\ref{fig:spin_ind}.  For $CP$-conserving neutralino 
interactions, these amplitudes were calculated long ago in~\cite{Goodman:1984dc,Griest:1988ma,
Srednicki:1989kj,Giudice:1988vs,Shifman}, and extended in 
\cite{Falk:1998xj} to include $CP$-violating effects.

In our conventions, the contributions to $C_{q_i}$ due to squark exchange in the 
$s$- and $u$-channels at zero momentum transfer read%
	\footnote{We have $i=1,2,3$ for generation indices and $s=1,\ldots,6$ for squark 
	mass eigenstates. To recover the expressions in~\cite{Falk:1998xj}, one needs 
	to make the identification $C^{\tilde{q}}_{u_1} \leftrightarrow \alpha^{}_{3_1}$ 
	etc.}
\begin{align}
	\bar{m}_{q_i}C^{\tilde{q}}_{q_i} &= \frac{1}{8} \sum_{s=1}^6 
	\left[ \frac{1}{(m_\chi+m_{q_i})^2-m_{\tilde{q}_s}^2+i\epsilon} 
	+ \frac{1}{(m_\chi-m_{q_i})^2-m_{\tilde{q}_s}^2+i\epsilon} \right] 
	\Re\{ \Gamma_L^{q_i\tilde{q}_s*}\Gamma_R^{q_i\tilde{q}_s} \}\,, 
	\label{eq:C sqrk}
\end{align}
where there is no sum over $i$, and a pole mass $m_{q_i}$ enters in the squark 
propagator.  Since we work at $M_\mathrm{SUSY}$, the running quark masses 
$\bar{m}_{q_i}$ must also be evaluated at this scale.  For the Higgs-exchange 
contribution we have%
	\footnote{In principle, the $CP$-odd Higgs can contribute to SI $\chi$--quark 
	scattering if $\mu$ or the Yukawa couplings $Y_{q_i}$ are allowed to be complex. 
	We exclude this possibility in our numerical analysis since we take real $\mu$ 
	and the real part of $Y_{q_i}$ after threshold corrections are included.}
\begin{equation}
	\bar{m}_{q_i}C_{q_i}^{h,H} = \frac{1}{2}\sum_{k=1}^2\frac{1}{m_{H_k}^2}
	\Re\{\Gamma_{\chi\chi}^{H_k}\}\Re\{\Gamma_{q_iq_i}^{H_k}\}\,, 
	\label{eq:C higgs}
\end{equation}
where ${H_1} \equiv H$ and ${H_2} \equiv h$. We assume that $m_h \simeq 125$ GeV 
is the mass of the Higgs-like resonance found at the LHC~\cite{Aad:2012tfa,Chatrchyan:2012ufa} 
and  $m_h<m_H$.  In general, the $\chi q_i\tilde{q}_s$ and $H_k^0\chi\chi$ couplings 
appearing in \eqref{eq:C sqrk} and \eqref{eq:C higgs} are complicated expressions 
involving the mixing matrices $Z^\chi$ and $Z^{\tilde{q}}$. Thus the SI 
cross section is typically determined numerically.  However, it is 
known~\cite{Arnowitt:1995vg,Hofer:2009xb,Crivellin:2011jt} that one can obtain 
analytic results by diagonalizing $M^\chi$ perturbatively%
	\footnote{For real $M_{1,2}$ and $\mu$, the exact diagonalization of $M^\chi$ 
	is known \cite{Guchait:1991ia,ElKheishen:1992yv,Barger:1993gh}, although 
	the resulting formulae are not simple.}
in powers of $v/M_\mathrm{SUSY}$.  For complex $M_{1,2}$ and $\mu$, one finds to 
leading order
\begin{align}
	Z^{\chi}_{11}	&=	e^{-\tfrac{i}{2}\phi_{M_1}} + \Order(v^2/M_\mathrm{SUSY}^2) \,, \notag \\
	Z^\chi_{21} 	&=	\Order(v^2/M_\mathrm{SUSY}^2) \,, \notag \\
	Z^\chi_{31} 	&=	-\frac{e^{-\tfrac{i}{2}\phi_{M_1}}}{\sqrt{2}} 
	\frac{g_1 v}{|M_1|^2-|\mu|^2} (M_1c_\beta + \mu^* s_\beta) 
	+ \Order(v^2/M_\mathrm{SUSY}^2)\,, \notag \\
	Z^\chi_{41}		&=	\frac{e^{-\tfrac{i}{2}\phi_{M_1}}}{\sqrt{2}} 
	\frac{g_1 v}{|M_1|^2-|\mu|^2} (M_1s_\beta + \mu^* c_\beta) 
	+ \Order(v^2/M_\mathrm{SUSY}^2)\,,
	\label{eq:Zchi}
\end{align}
where $\phi_{M_1}$ is the phase of $M_1$.  Note that the presence of a pole at 
$|M_1|=|\mu|$ has no physical meaning: it is a consequence of the fact that we 
assume a bino-like LSP and used non-degenerate perturbation theory to diagonalize 
the neutralino mass matrix. 

In Appendix~\ref{AppB}, we show how (\ref{eq:Zchi}) can be used to simplify 
(\ref{eq:C sqrk}) and (\ref{eq:C higgs}) if flavor-violating 
effects are neglected,%
	\footnote{Flavor violation in DM direct detection is strongly suppressed since 
	the effect can only enter via double flavor changes $f\to j\to f$ which are 
	experimentally known to be small.  Furthermore, the effect of flavor off-diagonal 
	entries can be largely absorbed by a change of the physical squark masses.}
while allowing for non-universal $A$-terms and squark masses. The resulting 
expressions read
\begin{align}
C_{u_i}^{\tilde{q}}	&= \frac{g_1^2}{8} \left[ \frac{2}{9} X_{u_i} L_{u_i}^+ R_{u_i}^+ 
										+\frac{1}{6} \frac{M_1+\mu\cot\beta}{M_1^2-\mu^2} 
										\left( L_{u_i}^+ - 4R_{u_i}^+  \right)\right]
										+ (L^+,R^+) \leftrightarrow (L^-,R^-)	\,, 
										\label{up-squark-contribution} \\
C^{\tilde{q}}_{d_i}	&= -\frac{g_1^2}{8} \left[ \frac{1}{9}X_{d_i} L_{d_i}^+ R_{d_i}^+ 
										+ \frac{1}{6} \frac{M_1+\mu\tan\beta}{M_1^2-\mu^2} 
										\left(L_{d_i}^+ + 2R_{d_i}^+ \right) \right]
										+ (L^+,R^+) \leftrightarrow (L^-,R^-)\,,	
										\label{down-squark-contribution} \\
C_{u_i}^{h,H}	&= \frac{g_1^2}{4}\frac{1}{M_1^2-\mu^2} \left[ 
							(M_1 + \mu \cot\beta) \left(\frac{c_\alpha^2}{m_h^2}
							+\frac{s_\alpha^2}{m_H^2} \right) - (M_1\cot\beta + \mu) 
							s_\alpha c_\alpha \left(\frac{1}{m_h^2} - \frac{1}{m_H^2} \right) 
							\right] \,,  \label{up-quark-Higgs-contribution}\\
C_{d_i}^{h,H}	&= \frac{g_1^2}{4}\frac{1}{M_1^2-\mu^2} \left[ (M_1 + \mu\tan\beta) 
							\left(\frac{s_\alpha^2}{m_h^2}+\frac{c_\alpha^2}{m_H^2} \right) 
							- (M_1\tan\beta + \mu) s_\alpha c_\alpha \left(\frac{1}{m_h^2} 
							- \frac{1}{m_H^2} \right) \right]  \label{eq:C Higgs dwn}\,,
\end{align}
where the squark mixing is defined as
\begin{equation}
	X_{u_i} \equiv A_u^{ii} + \mu\cot\beta \qquad \mbox{and} 
	\qquad X_{d_i} \equiv A_d^{ii} + \mu\tan\beta\,, \label{Xq}
\end{equation}
while the squark propagators are 
\begin{align}
	S^{\pm}_{q_i} = 
	\frac{1}{(m_\chi\pm m_{q_i})^2 - m_{\tilde{q}_i^S}^2 + i\epsilon}
	\qquad \mbox{for } S=L \mbox{ or } R\,, 
\end{align}
and $m_{\tilde{q}_i^L}^2$ and $m_{\tilde{q}_i^R}^2$ are the upper and lower 
diagonal components of the squark (mass)$^2$ matrices in (\ref{eq:sqrk mass}).  
In deriving (\ref{up-quark-Higgs-contribution}-\ref{eq:C Higgs dwn}), we have 
imposed $CP$ conservation so that the neutralino mass parameters $M_{1,2}$ and $\mu$ 
are real.  We also take $0<\beta<\pi/2$ and $M_{1,2}>0$ so that both signs of $\mu$ 
are allowed.  Expressions for $C_{q_i}$ in the $CP$-violating case are provided in Appendix~\ref{AppB}.

Note that:
\begin{enumerate}
\item The simplified expressions in~(\ref{up-squark-contribution}-\ref{down-squark-contribution}) 
are valid provided the squarks are sufficiently heavy, i.e.\ if 
$m_q^2 + m_\chi^2 \ll m_{\tilde{q}}^2$.  This requirement is not met for light 
third-generation squarks, and thus (\ref{up-squark-contribution}-\ref{down-squark-contribution}) 
must be corrected to account for the one-loop result~\cite{Drees:1993bu}.  To do so, 
we follow the prescription adopted in \cite{Belanger:2008sj} and replace all 
tree-level squark propagators
\begin{equation}
	S_{q_i}^{\pm} \to -K(\pm,m_{q_i},m_{\tilde{q}_i}^S,m_\chi)
\end{equation}
in terms of a linear combination $K$ of one-loop functions $I_i(m_\chi,m_q,m_{\tilde{q}})$,%
	\footnote{The term proportional to $I_3$ in~(A5) of~\cite{Belanger:2008sj} 
	is missing a factor of $m_q$.}
\begin{align}
	K(\alpha,m_q,m_{\tilde{q}},m_\chi) 
	&= \tfrac{3}{2}m_q \left[ m_q (I_1 - \tfrac{2}{3}m_\chi^2 I_3) - \alpha m_\chi 
	\left(I_2 - \tfrac{1}{3}I_5 - \tfrac{2}{3}m_\chi^2 I_4 \right) \right]\,,
	\label{K function}
\end{align}
whose form is given in Appendix B of~\cite{Drees:1993bu}.  In the heavy squark 
limit, the function $K$ agrees with $S_{q_i}^{\pm}$ to leading order in 
$m_{\tilde{q}}^{-2}$.
\item We have made use of the tree-level relation $Y_{q_i} = \bar{m}_{q_i} / v_q$ 
in order to obtain (\ref{up-squark-contribution}-\ref{eq:C Higgs dwn}).  
For down quarks, however, this relation can be modified by one-loop graphs which 
induce an effective coupling between $d_i$ and the neutral component of $H_u$.  
These corrections~\cite{Hall:1993gn,Carena:1994bv,Carena:1999py,Hofer:2009xb,Crivellin:2011jt} 
are non-decoupling and enhanced by a factor of $\tan\beta$.%
	\footnote{In principle, large $A$-terms can also change the values of $Y_{q_i}$ 
	significantly~\cite{Banks:1987iu,Borzumati:1999sp,Crivellin:2010gw,Crivellin:2010er,Crivellin:2011jt}. However, this effect drops out in the Higgs--quark--quark 
	couplings where the effective (physical) mass enters. Furthermore, since we assume 
	flavor-universal $A$-terms in our numerical analysis, the effect cannot be very 
	large without violating vacuum stability bounds.}
For example, the gluino contribution at one-loop modifies the tree-level relation 
so that 
\begin{equation}
	Y_{d_i} = \frac{\bar{m}_{d_i}}{v_d(1+\epsilon_i \tan\beta)}\,,
	\label{Yd thresh}
\end{equation}
where $\epsilon_i \simeq - \frac{2\alpha_s}{3\pi}m_{\tilde{g}} \mu^* C_0(m_{\tilde{g}}^2,m_{\tilde{d}^L_i}^2,m_{\tilde{d}^R_i}^2)$ 
and 
\begin{equation}
	C_0(a^2,b^2,c^2) = \frac{b^2}{(a^2-b^2)(c^2-b^2)} \log \frac{a^2}{b^2} 
	+ \frac{c^2}{(a^2-c^2)(b^2-c^2)} \log \frac{a^2}{c^2}\,.
\end{equation}
Since $\bar{m}_{q_i}C_{q_i}$ is proportional to $Y_{q_i}v_q$, we can account for 
(\ref{Yd thresh}) by a simple rescaling of the Wilson coefficients
\begin{equation}
	C_{d_i}^{\tilde{q},H} \rightarrow 
	\frac{v_d Y_{d_i}}{\bar{m}_{d_i}} C_{d_i}^{\tilde{q},H}\,,
	\label{eq:threshold}
\end{equation}
where we include corrections~\cite{Hall:1993gn,Carena:1994bv,Carena:1999py,Hofer:2009xb,Crivellin:2011jt} 
beyond the gluino loop (\ref{Yd thresh}).  These threshold corrections feature in 
our analysis of heavy Higgs $H$ and sbottom contributions 
(Sec.\ \ref{sec:simplified models}) to the SI amplitude.  Note that corrections 
to the light Higgs coupling $h\bar{d}d$ cancel in the relation 
$\bar{m}_{u_i} = Y_{u_i}v_u$.
\end{enumerate}

\section{Simplified models: blind spots and isospin violation}
\label{sec:simplified models}
\begin{figure}[t]
	\centering\includegraphics[scale=0.8]{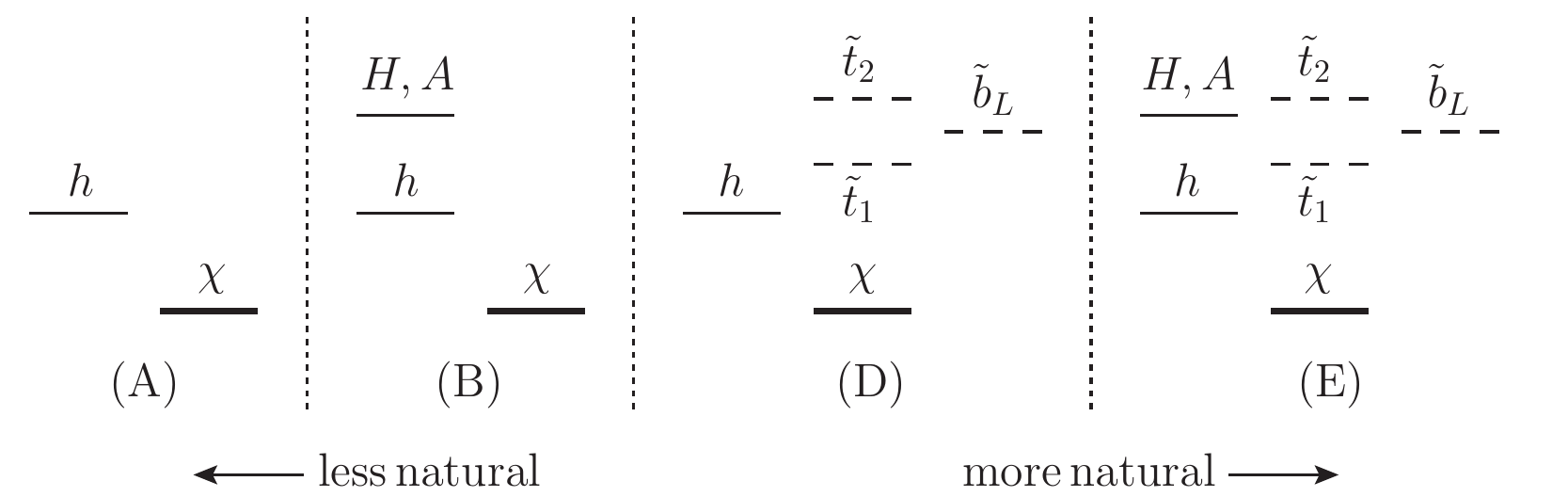}
	\caption{Spectra of the simplified models (A-D) 
	considered in this work.  For each model, the LSP $\chi$ is assumed to be bino-like  
	and may be accompanied by a nearly degenerate tau slepton in order to produce 
	the observed DM relic density.  The SM-like Higgs is denoted by $h$, 
	while all other states are assumed to lie below 1 TeV, including Higgsinos 
	(not shown). From left-to-right, the spectra become increasingly more natural 
	as one includes the additional Higgs states $H,A$ and third-generation squarks 
	$\tilde{t}_1,\tilde{t}_2,\tilde{b}_L$.  In general, $\chi$ may be heavier than 
	$h$, $H$, and $A$.}
	\label{fig:simple}
\end{figure}

We now apply our analytic results (\ref{up-squark-contribution}-\ref{eq:C Higgs dwn}) 
to four simplified models; each motivated by the following experimental and naturalness considerations. 
Firstly, the ATLAS~\cite{Aad:2012tfa} and CMS~\cite{Chatrchyan:2012ufa} experiments 
at the LHC have discovered a Higgs boson with SM-like properties and a mass below 
the upper bound $\lesssim $~135 GeV of the MSSM.  Secondly, a natural resolution 
of the gauge hierarchy problem requires several 
conditions~\cite{Barbieri:1987fn,Dimopoulos:1995mi,Cohen:1996vb} to be met:
\begin{itemize}
	\item In order to cancel the top-quark correction to the Higgs mass parameter 
	$m_{H_u}^2$, top squarks must be light with masses in the sub-TeV range;
	\item The gluino mass must be around a TeV in order to prevent radiative 
	corrections driving the stop masses too heavy;
	\item Light Higgsinos must be present in the spectrum so that tree-level 
	electroweak symmetry breaking implies that $\mu \sim v$ is satisfied.
\end{itemize}
It has also been observed~\cite{Katz:2014mba} that naturalness constrains the 
additional Higgs bosons $H,H^{\pm},A$ to not be too heavy.  Barring the gluino, 
the current experimental bounds on the masses of the above particles are rather 
weak.  In contrast, the mass of the gluino and squarks of the first two generations 
are constrained to lie above 1~TeV. Therefore, naturalness prompts us to consider 
the simplified models shown in Fig.\ \ref{fig:simple}, where we start from a 
minimal, light particle spectrum necessary to have bino-like DM scattering 
[model (A)] and successively include as active degrees of freedom those particles 
which are (a) required to be light by naturalness, and (b) relevant for DM direct detection. 
Note that due to $SU(2)_L$ invariance, the models (C-D) involving two light stops 
always require a light sbottom in the spectrum.  (Only if there is a single, mostly 
right-handed stop, can sbottoms be decoupled.) 

In general, a bino-like LSP produces a DM relic density that is too large in most of the parameter space considered in Sec.~\ref{sec:simplified models}.  However, the overproduction of bino-like DM in the MSSM can be diluted by either $s$-channel resonance exchange involving $Z,h,H,A$, or $\chi$--$\tilde{f}$ co-annihilation with a sfermion $\tilde{f}$ that is nearly degenerate in mass with $\chi$.%
	\footnote{See e.g.~\cite{Han:2013gba} for a detailed analysis of these effects in the pMSSM.}
Both mechanisms~\cite{Griest:1990kh} increase the annihilation cross section before thermal freeze-out and can produce the observed relic abundance.  In each of the models shown in Fig.\ \ref{fig:simple}, the relic density constraint may be satisfied by either mechanism or, if necessary, by  extending the spectrum to include a tau slepton $\tilde{\tau}$ which generates additional co-annihilations~\cite{Ellis:1998kh,Ellis:1999mm}.  Since the $\tilde{\tau}$ mass can be tuned without affecting naturalness or DM direct detection, we do not consider the DM relic density constraint in our subsequent analysis. 

Similarly, we do not consider the constraint from the anomalous magnetic moment of the muon $a_\mu$, whose world average is dominated by the Brookhaven measurement~\cite{Bennett:2006fi}. 
The resulting value deviates from the SM prediction by $3$--$4\sigma$, depending on the details of
the evaluation of the hadronic contributions~\cite{Jegerlehner:2009ry,Prades:2009tw}.
Recent developments in the evaluation of the SM prediction include: 
the QED calculation has been carried out at $5$-loop accuracy~\cite{Aoyama:2012wk}, 
after the Higgs discovery~\cite{Aad:2012tfa,Chatrchyan:2012ufa}
the electroweak contribution is complete at two-loop order~\cite{Gnendiger:2013pva,Czarnecki:2002nt},
and hadronic corrections have been considered at third order in the fine-structure constant~\cite{Kurz:2014wya,Colangelo:2014qya}. 
Although an improved determination of the leading hadronic contribution, hadronic vacuum polarization, mainly requires
improved data input, see~\cite{Jegerlehner:2009ry,Davier:2010nc,Hagiwara:2011af,Blum:2013xva}, the uncertainties in the subleading hadronic-light-by-light contribution have been notoriously difficult to estimate due to substantial model dependence~\cite{Jegerlehner:2009ry,Prades:2009tw,Benayoun:2014tra}. Recently, data-driven techniques have been put forward to reduce the model dependence based on dispersion relations~\cite{Colangelo:2014dfa,Colangelo:2014pva,Pauk:2014rfa,Colangelo:2015ama}, and a first lattice calculation has become available~\cite{Blum:2014oka}.  
All these efforts are motivated by two new experiments, at FNAL~\cite{Grange:2015fou} and J-PARC~\cite{Saito:2012zz}, which each aim at improving the measurement by a factor of $4$ and thus help clarify the origin of the discrepancy between experiment and the SM prediction.

Should the discrepancy persist, an explanation within the MSSM is possible provided certain assumptions are made about the SUSY parameters entering the smuon, chargino, and neutralino mass matrices.  If these parameters are all equal to $M_\mathrm{SUSY}$, then a positive contribution to $a_\mu$ requires $\mathrm{sign}(\mu M_2)>0$ since the dominant one-loop amplitude scales approximately with $\mu M_2 \tan\beta / M_\mathrm{SUSY}^2$; see e.g.\ the review~\cite{Stockinger:2006zn} and references therein.  In the blind spot regions where $\mu <0$, this condition would require us to relax the assumption that $M_2>0$.  However, the requirement $\mathrm{sign}(\mu M_2) >0 $ does not necessarily apply if the SUSY mass parameters are non-degenerate.  For example, it has been shown~\cite{Cho:2011rk,Fargnoli:2013zda,Fargnoli:2013zia} that a positive contribution to $a_\mu$ can arise if $|M_1|, m_{\tilde{\mu}_R} \ll |M_2|, m_{\tilde{\mu}_L}$, in which case $\mu$ and $M_{1,2}$ must have \textit{opposite} sign.  The key point is that neither the sign of $M_2$ nor the smuon masses are relevant for our analysis of SI scattering, so it would be possible to account for the experimental value of $a_\mu$ by a suitable choice of these parameters.  Furthermore, the discrepancy could also be explained by large $A_\mu$ terms~\cite{Borzumati:1999sp,Crivellin:2010ty,Endo:2013lva} not correlated with DM scattering.  

We conclude this section by anticipating a key result of our analysis: 
isospin-violating effects can be magnified in the proximity of blind spots, where 
the SI direct-detection cross section lies below the lower bounds set by the 
irreducible neutrino background. For these parameter-space configurations, the 
SI amplitude itself becomes tiny and hence more susceptible to small variations 
in the input quantities, such as corrections from isospin breaking. In particular,
the ratio of proton and neutron SI cross sections becomes very sensitive to the 
values of the scalar matrix elements and their uncertainties $\delta f_{n,p}$,
\begin{equation}
	\left(\delta\frac{f_n}{f_p}\right)^2 = \left(\frac{\delta f_n}{f_p}\right)^2 
	+ \left(\frac{f_n\delta f_p}{f_p^2}\right)^2\,,
	\label{eq:fNofP uncertainty}
\end{equation}
so that the overall uncertainty on $f_n/f_p$ can become large near blind spots 
where $f_p\simeq 0$.  In each of the four simplified models (A-D), we examine   
the amount of isospin violation associated with the three methods of 
Sec.~\ref{sec:hadronic}.

\subsection{SM-like Higgs exchange}
%
\begin{figure}[t]
        \centering
        \subfloat{\includegraphics[scale=0.6]{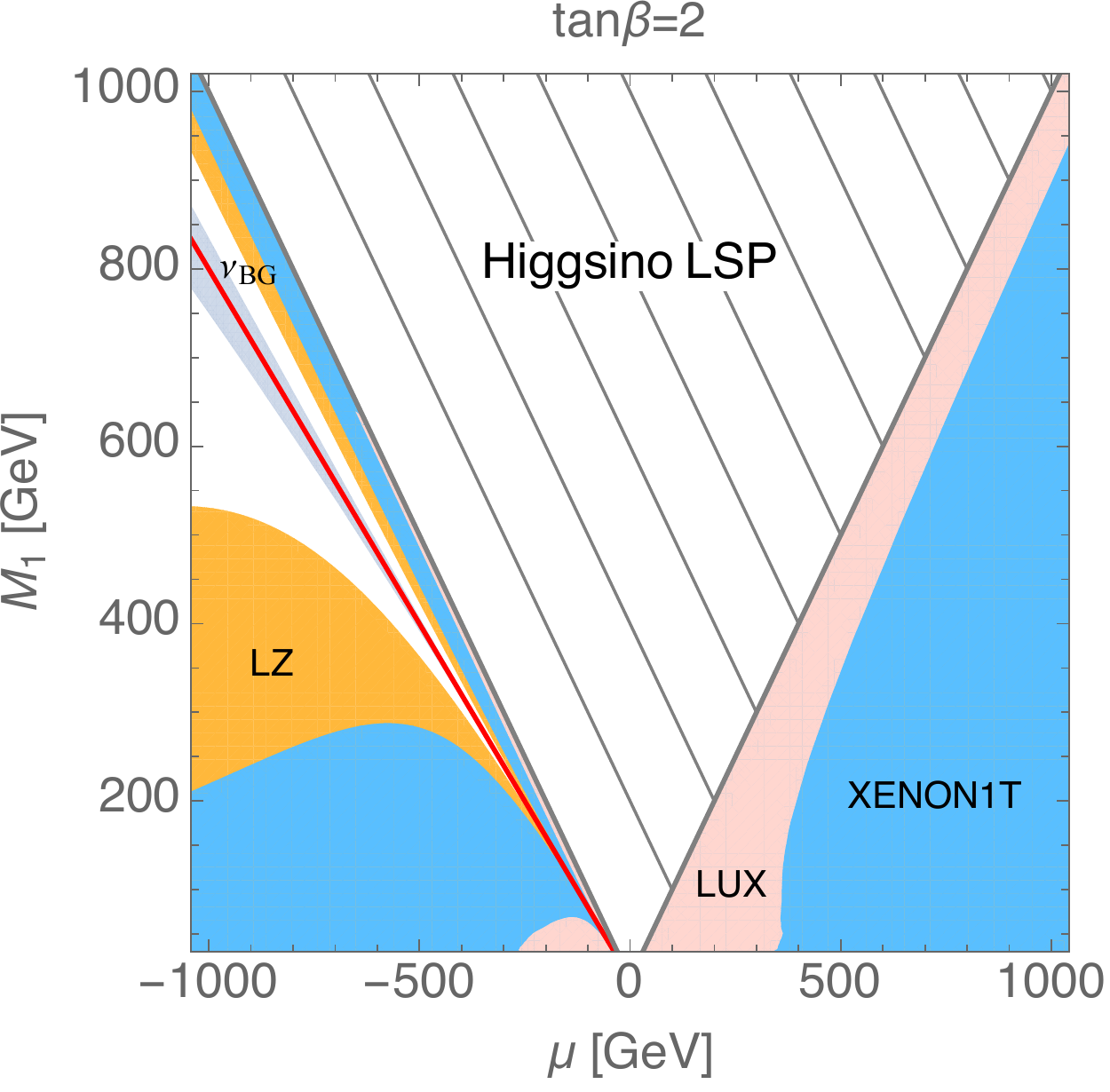}}
        \qquad
        \centering
        \subfloat{\includegraphics[scale=0.6]{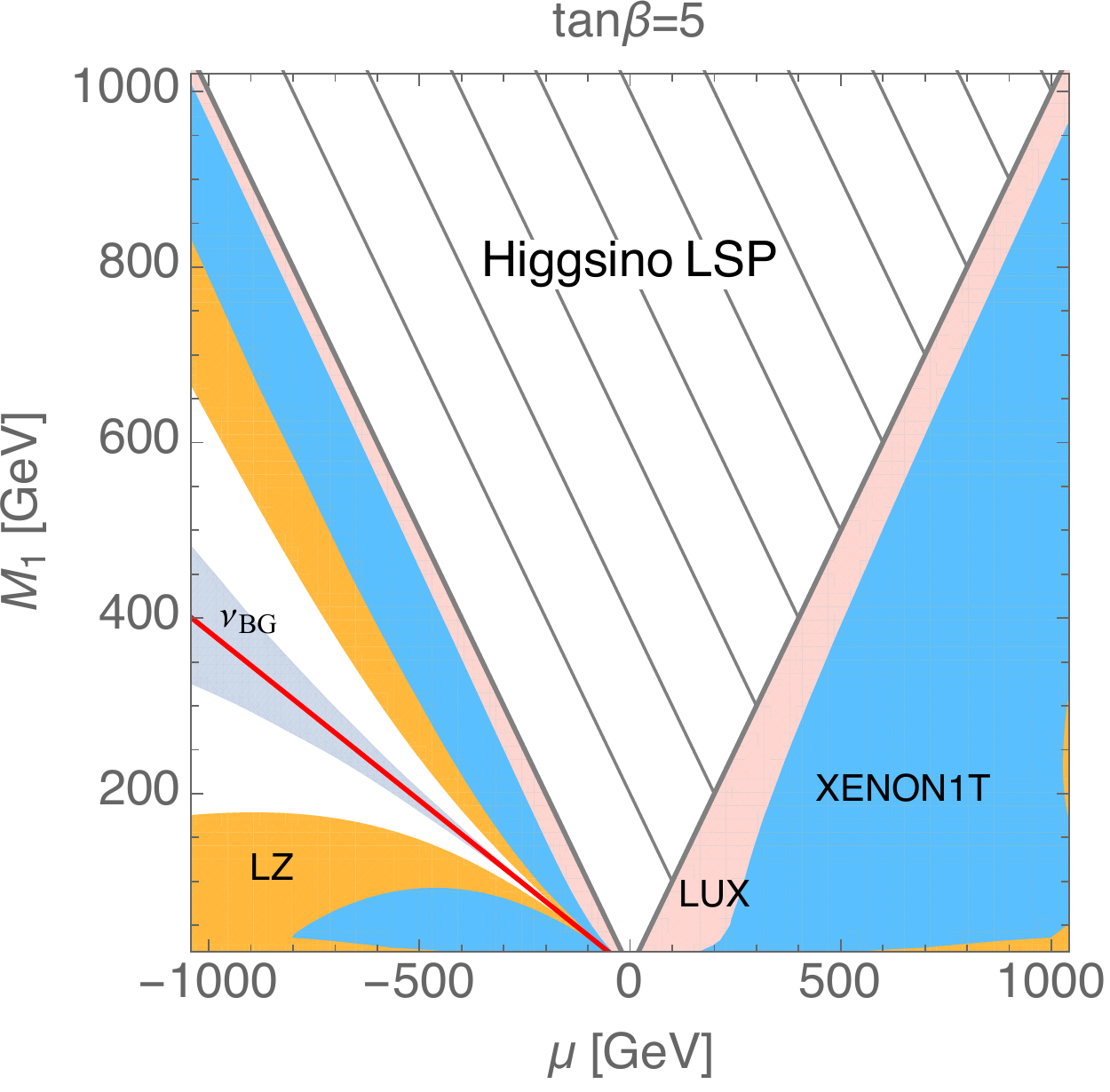}}
        \qquad
        \centering
        \subfloat{\includegraphics[scale=0.6]{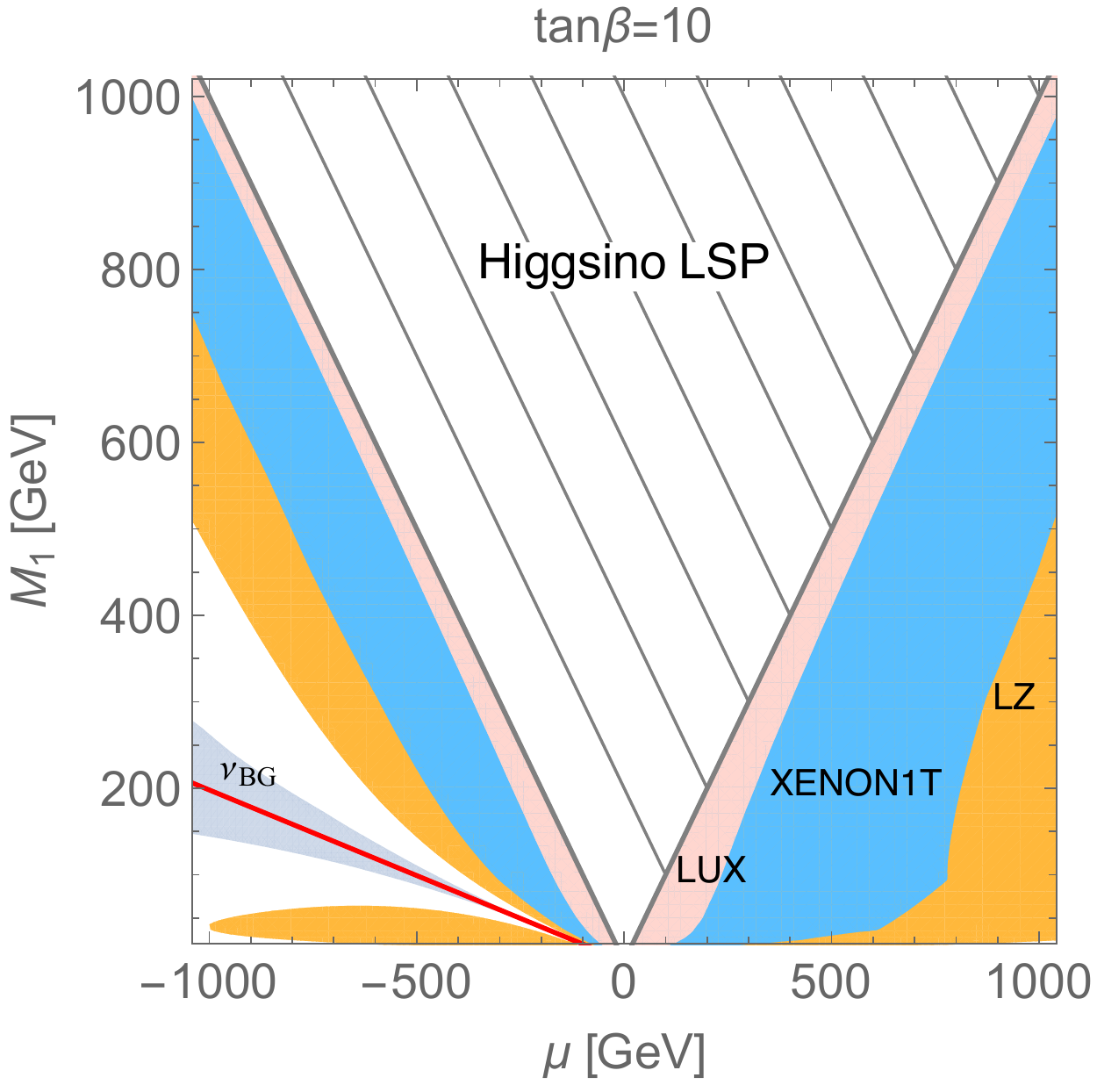}}
        \qquad
        \centering
        \subfloat{\includegraphics[scale=0.6]{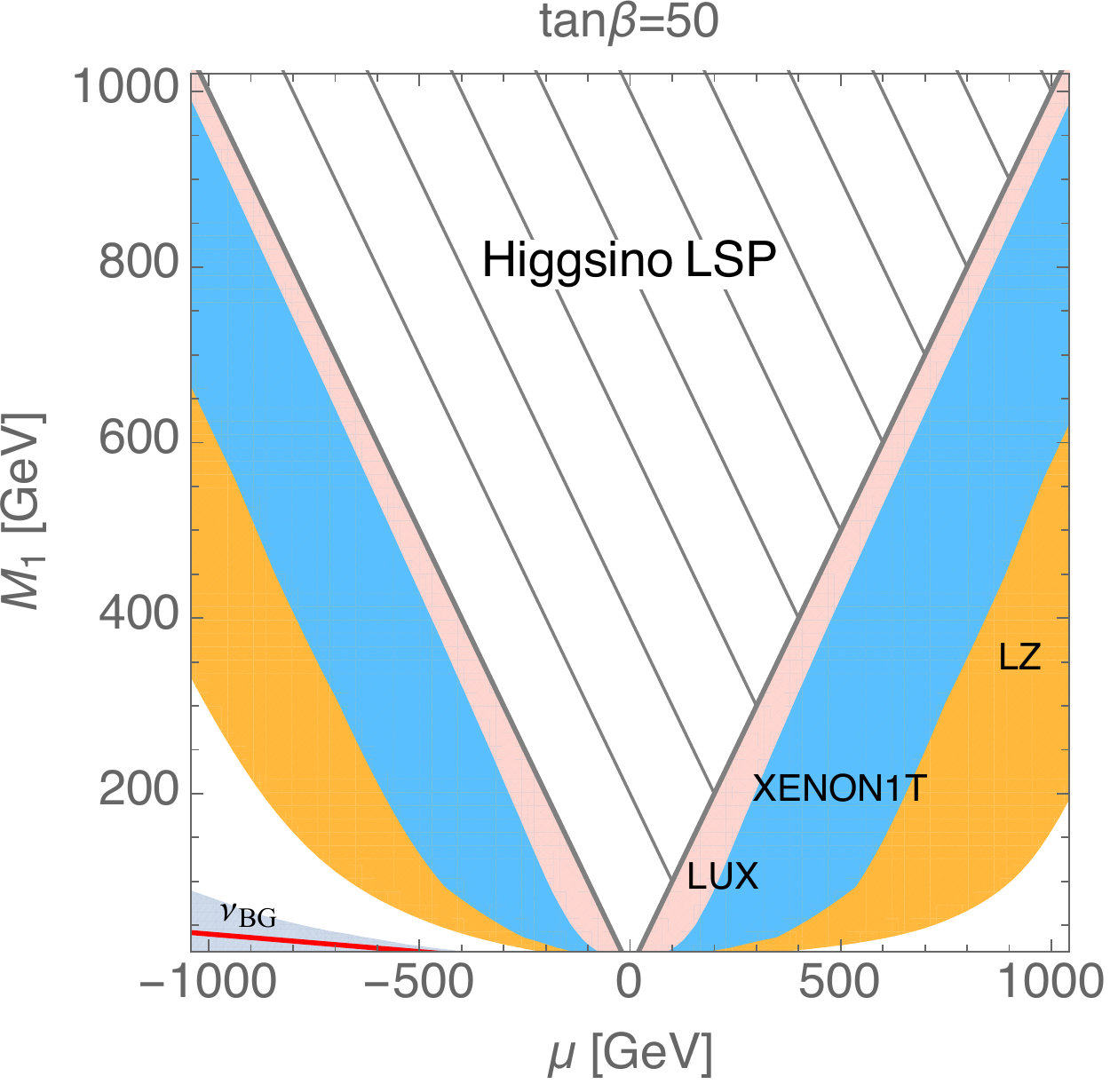}}
\caption{Current and projected limits on SI $\chi$--xenon scattering due to $h$ 
exchange with $\tan\beta=2$ (top left), 5 (top right), 10 (bottom left), and 50 (bottom right). The pink band shows the existing constraints 
from LUX~\cite{Akerib:2013tjd}, while projected limits from XENON1T~\cite{Aprile:2012zx} 
and LZ \cite{Malling:2011va} are given by the blue and orange regions respectively.  
The blind spot where the SI cross section vanishes is denoted by the red line and 
lies within the irreducible neutrino background ($\nu_\mathrm{BG}$) shown in gray.  We assume the LSP is bino-like (Fig.\ \ref{fig:simple}), so do not consider the triangular, hatched region where $\mu < M_1$ and $\chi$ becomes Higgsino-like.}
	\label{fig:m1mu}
\end{figure}
We begin by considering the minimal particle spectrum for which an observable 
SI cross section is possible.  From the rightmost diagram in Fig.\ \ref{fig:spin_ind}, 
it seems reasonable to conclude that the SM-like Higgs and the bino-like LSP is sufficient 
in this case. However, in the limit $m_A \gg m_Z$, 
(\ref{up-quark-Higgs-contribution}) and (\ref{eq:C Higgs dwn}) become
\begin{equation}
	C_{u_i}^{h} = C_{d_i}^h = 
	\frac{g_1^2}{4m_h^2}\frac{1}{M_1^2-\mu^2} (M_1 + \mu s_{2\beta}) \,, 
	\qquad u_i = u,c,t, \qquad d_i = d,s,b\,,
	\label{eq:rh} 
\end{equation}
and thus the scattering amplitude decouples with the Higgsino mass $\mu$.  It 
follows that a measurable cross section due to Higgs exchange implies the 
presence of light Higgsinos in the spectrum, thereby satisfying one of the minimal 
naturalness requirements.  Although this feature does not prevent the reintroduction 
of fine-tuning in the MSSM altogether, it becomes relevant in our subsequent analysis 
where light stops are added to the spectrum.

To compare (\ref{eq:rh}) to data, we first note that $C_{q_i}^h$ vanishes when
\begin{equation}
M_1 + \mu s_{2\beta} = 0\,,
\label{eq:h blind}
\end{equation}
and thus a blind spot arises in the SI cross section provided $\mu$ is negative.  
The prospects for constraining this feature (\ref{eq:h blind}) 
have been extensively analyzed~\cite{Cheung:2012qy} for $\chi$--nucleon scattering. 
To examine isospin violation, however, we need limits on $\chi$--{\it nucleus} 
cross sections, so we use (\ref{eqn:cross SI}) in order to constrain the relevant 
parameter space.

Let us first consider the limits associated with (\ref{eq:rh}) when the scalar 
matrix elements $f_q^N$ of \sone~are employed.  In Fig.\ \ref{fig:m1mu}, we update 
the results from~\cite{Cheung:2012qy} and show constraints for various values of 
$\tan\beta$ in the $(\mu,M_1)$ plane from current and upcoming xenon experiments.
For $\mu>0$, we find that only a narrow strip is excluded by the existing limits from LUX~\cite{Akerib:2013tjd}, while the projected reach from XENON1T~\cite{Aprile:2012zx} and LZ~\cite{Malling:2011va} will probe most of the naturalness-preferred region where $\mu$ is of order $v$.  As $\tan\beta$ is increased, the term $\propto \mu$ in $C_{q_i}^h$ is suppressed, thereby weakening the direct detection limits. If no signal is seen at LZ, then the allowed parameter space is focused towards $\tan\beta=50$ and values of $M_1 \lesssim 200$ GeV.
In the $\mu < 0$ region and for small $\tan\beta$, the naturalness-preferred values of $\mu$ occur at $|\mu| \simeq M_1$ and are  concentrated near the blind spot.  Although the irreducible neutrino background make this region difficult to probe experimentally, larger values of $\tan\beta$ decrease the blind spot slope, so that natural values of $\mu$ become allowed for $M_1 \lesssim 300$-$400$ GeV.

\begin{figure}[t]
	\centering\includegraphics[scale=0.85]{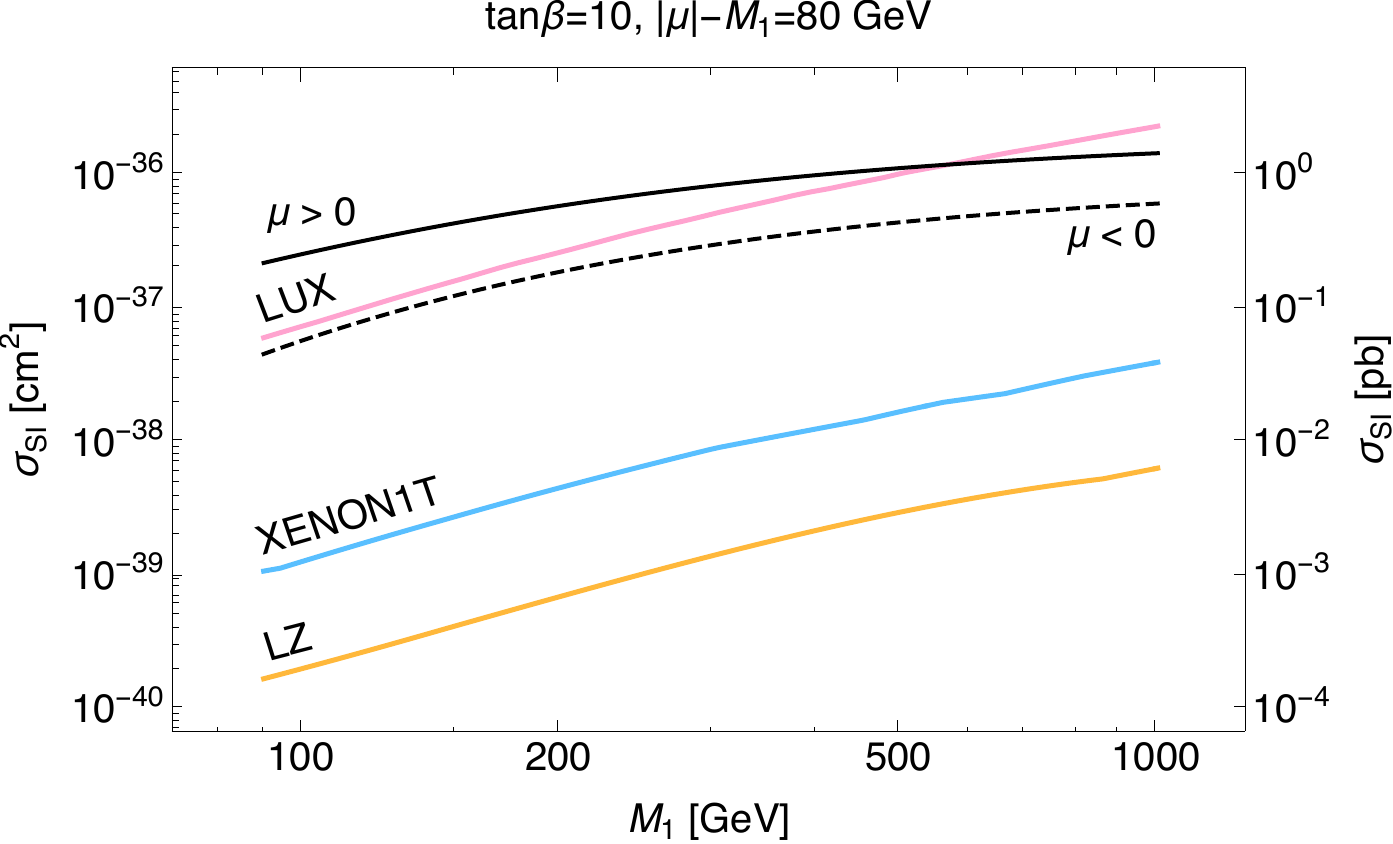}
	\caption{SI $\chi$--xenon cross sections for $h$ exchange with $\tan\beta = 10$ 
	and a small mass difference $|\mu|-M_1=80$ GeV between the Higgsino and bino.  
	The solid (dashed) black line corresponds to $\mu>0$ ($\mu<0$).  Shown are the 
	limits from LUX~\cite{Akerib:2013tjd},  XENON1T~\cite{Aprile:2012zx}, and 
	LZ~\cite{Malling:2011va}, with the same color coding as in Fig.\ \ref{fig:m1mu}.}
	\label{fig:cross_higgs}
\end{figure}
By taking a slice through the $(\mu,M_1)$ plane, we can also extract the limits 
due to a small mass splitting $|\mu|-M_1 = 80$ GeV between the bino and 
Higgsinos.  
This choice is motivated by the current CMS results~\cite{CMS:2014jfa} on same-flavor opposite-sign dilepton	searches. Here CMS sees a $2.6\sigma$ deviation which can be explained by a heavier neutralino decaying to a lighter one.
Fig.\ \ref{fig:cross_higgs} shows the resulting constraints, where we plot the 
SI cross sections as a function of the bino mass.  For $\mu > 0$, the limits from 
LUX are stringent, with values below $M_1 \simeq 600$ GeV excluded.  The strength 
of these limits is due to an enhancement in the amplitude (\ref{eq:rh}) from both 
a nearly degenerate denominator and lack of interference in the numerator terms.  
For $\mu < 0$, there are no constraints from LUX, although XENON1T and LZ will 
exclude the whole parameter space in the absence of a DM signal.

\begin{figure}[t]
	\centering\includegraphics[scale=0.8]{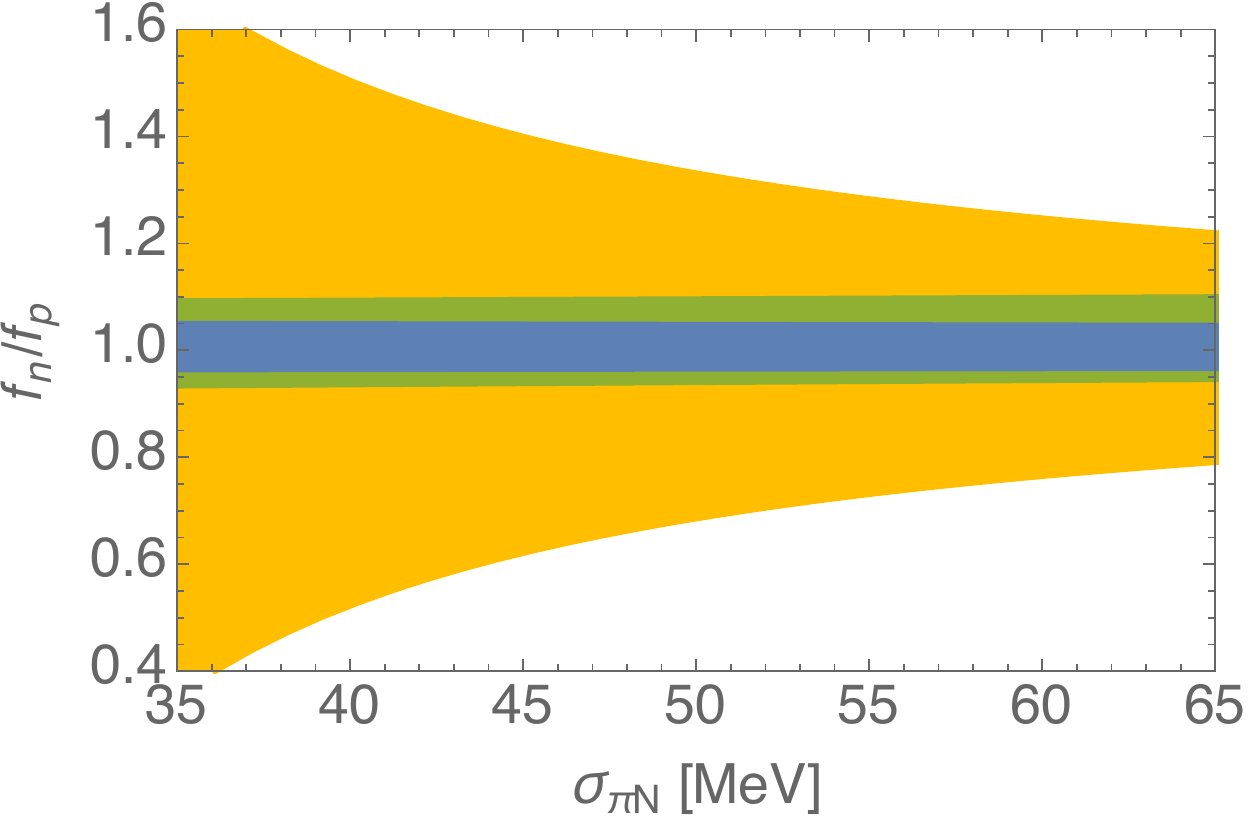}
	\caption{Amount of isospin violation in terms of $f_n/f_p$ due to $h$ exchange 
	in $\chi$--nucleus scattering.  The colored bands correspond to the $1\sigma$ 
	uncertainties associated with the different determinations of the scalar matrix elements 
	$f_q^N$ discussed in Sec.~\ref{sec:hadronic}.  The blue band corresponds to~\sone, 
	while the orange and green bands correspond to~\stwo\ and~\sthree\ respectively.}
	\label{fig:hNohP}
\end{figure}
We now examine the hadronic uncertainties associated with each of the three 
methods discussed in Sec.\ \ref{sec:hadronic}.  For $h$ exchange, the Wilson 
coefficient (\ref{eq:rh}) is independent of quark flavor, so the SI amplitude 
(\ref{fN}) factorizes 
\begin{align}
	\frac{f_N}{m_N} = C_{q_i}^{h} 
	\bigg( \frac{2}{9} + \frac{7}{9}\sum_{q=u,d,s} f_q^N \bigg)\,.
	\label{SM Higgs fN}
\end{align}
Evidently, the resulting SI cross section is sensitive to the value of $f_s^N$, 
with a dramatic effect observed~\cite{Giedt:2009mr} on the regions of excluded 
parameter space when the typically large value $f_s^N \approx 0.25$ of 
\sthree~is replaced with much smaller determinations (\ref{fs_lattice}) from the 
lattice.  We emphasize that this sensitivity is also present in {\it any} analysis 
of isospin violation, where $f_n/f_p$ is the quantity of interest.  For the present 
discussion, (\ref{SM Higgs fN}) implies that the ratio
\begin{equation}
	\frac{f_n}{f_p} = \left(\frac{m_n}{m_p}\right) 
	\frac{2+7\sum f_q^n}{2+7\sum f_q^p} 
	\label{eq:h ratio}
\end{equation}
is independent of $C_{q_i}^h$, and thus isospin violation is entirely determined 
by hadronic quantities. In Fig.\ \ref{fig:hNohP} we compare the uncertainties on 
$f_n/f_p$ as a function of $\sigma_{\pi N}$.  For Methods 1 and 3, we find stability 
across a large range of $\sigma_{\pi N}$ values, with isospin violation allowed 
at around the five and ten percent level respectively.  As noted in~\cite{Crivellin:2014cta}, 
this stability is due to the fact that the constant term of $\frac{2}{9}$ in 
(\ref{SM Higgs fN}) dominates the remainder whenever $f_s^N$ is fixed by lattice 
input.  In contrast, the $\chi$PT$_3$ formalism of \stwo~produces a strong dependence 
of $f_s^N$ on $\sigma_{\pi N}$, which in turn affects $f_n/f_p$.  From Fig.\ \ref{fig:hNohP}, 
isospin violation greater than $50\%$ is allowed, in marked contrast to the 
precision of \sone.  This example clearly demonstrates the huge uncertainties 
associated with \stwo, which, however, is still used in the 
literature~\cite{Buchmueller:2013rsa,Buchmueller:2014yva}.

\subsection{Light and heavy Higgs exchange}
\label{sec:h H exchange}
Let us now extend model (A) to include the heavy Higgs bosons $H,A,H^{\pm}$ 
[model (B) in Fig.\ \ref{fig:simple}].  The inclusion of these additional degrees 
of freedom is motivated by naturalness~\cite{Katz:2014mba}, however, only $H$ 
contributes to the SI cross section (Fig.\ \ref{fig:spin_ind}). 

From our simplified expressions (\ref{up-quark-Higgs-contribution}-\ref{eq:C Higgs dwn}), 
we see that the couplings to up and down quarks differ by a factor of $\tan\beta$, 
but are identical%
	\footnote{Up to threshold corrections (\ref{eq:threshold}), which enhance 
	$C_{d_i}^H$ by tens of percent at large $\tan\beta$.  Their inclusion does not 
	have a large impact on the numerical analysis.} 
among different generations $i=1,2,3$.  As a result, the SI amplitude may be 
expressed as
\begin{equation}
	\frac{f_N}{m_N} = C_{u_i}^{h,H} U_N + C_{d_i}^{h,H} D_N\,,
	\label{eq:h H amp}
\end{equation}
where 
\begin{align}
	U_N = f_u^N + 2f_Q^N \qquad \mbox{and} \qquad D_N = f_d^N + f_s^N + f_Q 
\end{align}
collect the scalar coefficients associated with the up- and down-type Wilson 
coefficients.  A blind spot occurs if the condition
\begin{equation}
	C_{u_i}^{h,H} U_N + C_{d_i}^{h,H} D_N \simeq 0
	\label{eq:h H blind}
\end{equation}
is satisfied, and the resulting suppression of the SI cross section has been 
identified numerically~\cite{Ellis:2000ds,Ellis:2000jd,Baer:2006te,Anandakrishnan:2014fia} 
and further studied analytically~\cite{Huang:2014xua}.  In the latter case, an 
explicit formula~\cite{Huang:2014xua} for the blind spot can be found for moderate 
to large values of $\tan\beta$ and $m_A>m_h$:
\begin{equation}
	\frac{2}{m_h^2} (M_1 + \mu s_{2\beta}) + \mu \tan\beta \frac{1}{m_H^2} \simeq 0\,.
	\label{eq:h H blind analytic}
\end{equation}
In effect, (\ref{eq:h H blind}) has been recast as an {\it interference}  
condition between the $h$ and $H$ amplitudes; a feature which has important 
consequences for isospin violation in the MSSM.  As with $h$ exchange, negative 
values of $\mu$ are required in order to generate the blind spot.  However, note 
that in the vicinity of (\ref{eq:h blind}), the first term in 
(\ref{eq:h H blind analytic}) is suppressed, so in some cases the contribution 
from $H$ exchange may dominate the scattering amplitude~\cite{Huang:2014xua}.
\begin{figure}[t]
	\centering\includegraphics[scale=0.8]{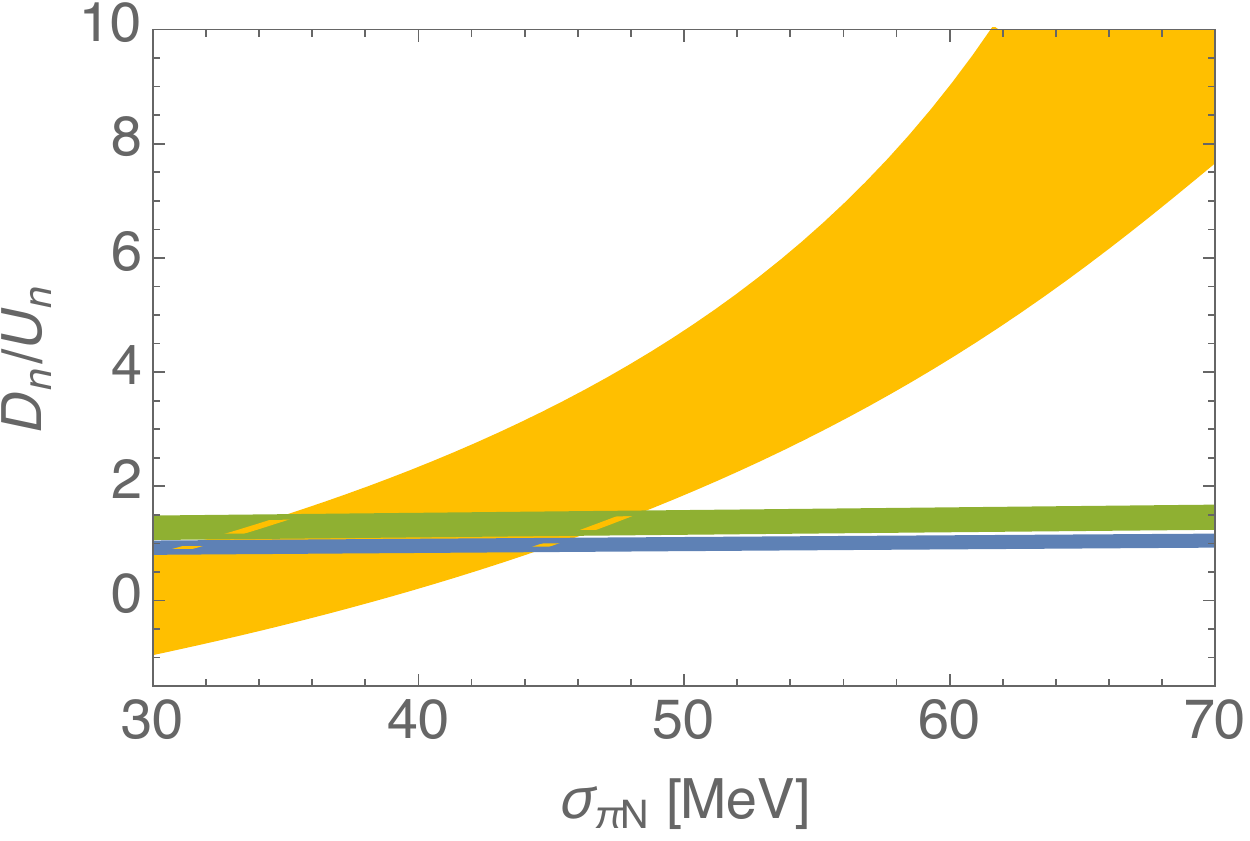}
	\caption{Dependence on $\sigma_{\pi N}$ in the ratio of the hadronic terms $U_N$ 
	and $D_N$ pre-multiplying the up- and down-type Wilson coefficient for $h,H$ 
	exchange (\ref{eq:h H amp}). Shown is the case for $N=$ neutron, with similar 
	results obtained for $N=$ proton. Color coding as in Fig.\ \ref{fig:hNohP}.}
	\label{fig:DNoUN}
\end{figure}

A crucial step in the derivation~\cite{Huang:2014xua} of (\ref{eq:h H blind analytic}) 
is the observation that $U_N \approx D_N$ numerically.  Deviations of $D_N/U_N$ 
from unity have the effect of shifting the location of the blind spot 
(\ref{eq:h H blind analytic}), so it is necessary to determine this ratio 
precisely.  In Fig.\ \ref{fig:DNoUN}, we display the sensitivity of $D_N/U_N$ to 
$\sigma_{\pi N}$ for each of the three methods of Sec.~\ref{sec:hadronic}.  Similar to 
our analysis of $h$ exchange (Fig.\ \ref{fig:hNohP}), we find that Methods 1 and 
3 are stable across a large range of $\sigma_{\pi N}$ values, with 
$D_N/U_N\simeq 1$ tightly constrained.  In contrast, \stwo~exhibits a strong 
dependence on $\sigma_{\pi N}$ and for $\sigma_{\pi N} \gtrsim45 $~MeV, the 
location of the blind spot (\ref{eq:h H blind analytic}) can get shifted by a factor 
of eight or more. These findings illustrate again the importance of using 
a well-controlled framework for the hadronic input quantities.

\begin{figure}[t]
	\centering
    \subfloat{\includegraphics[scale=0.8]{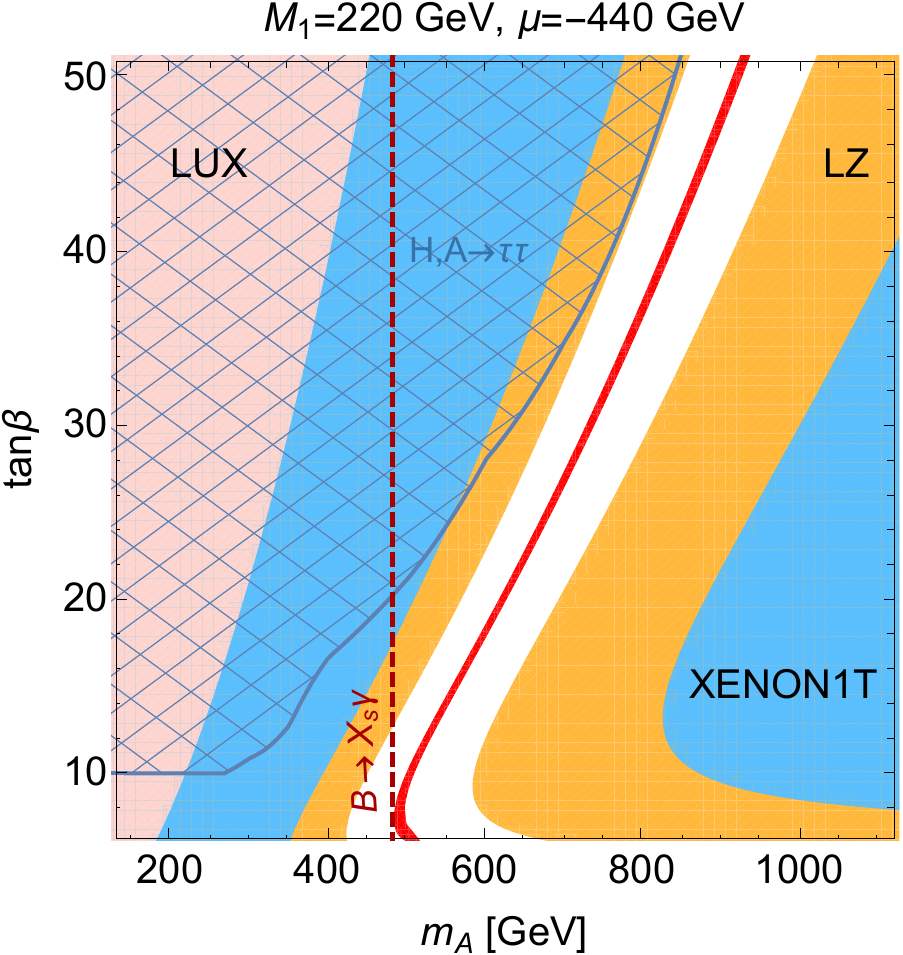}}
    \qquad
    \subfloat{\includegraphics[scale=0.8]{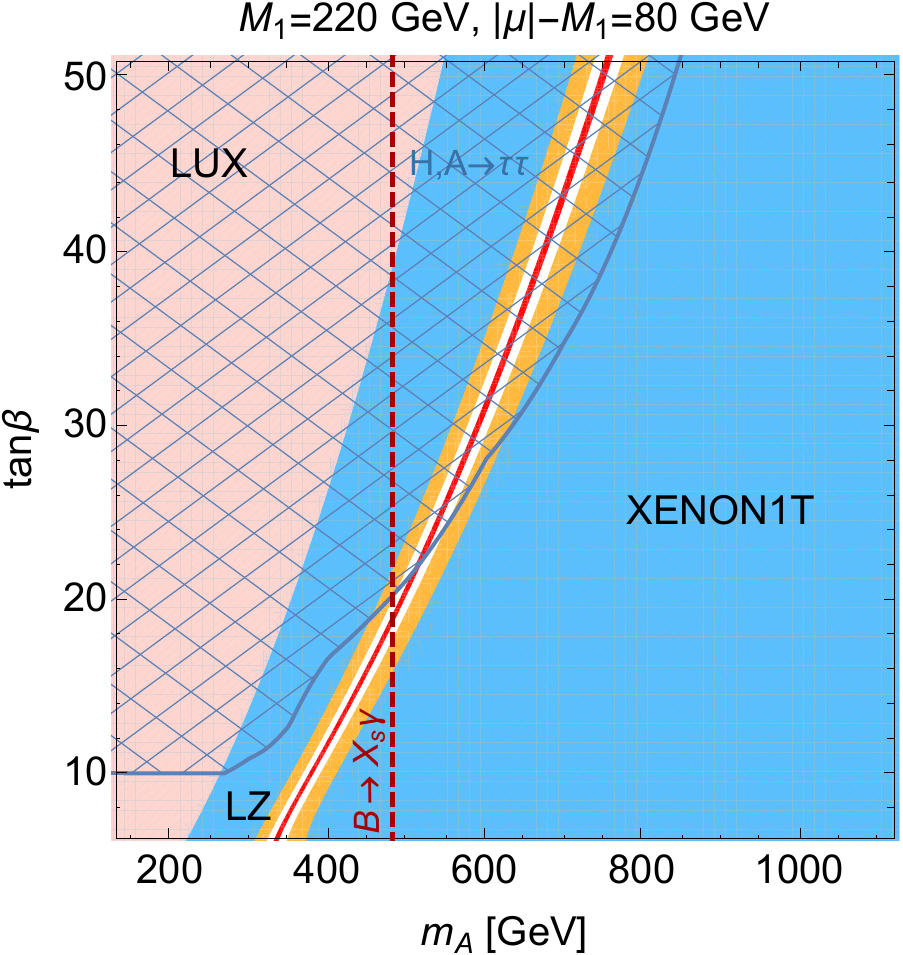}}
	\caption{Current and projected limits on SI $\chi$--xenon scattering due to $h,H$ 
	exchange with different benchmark values for $M_1$ and $\mu$.  Excluded regions 
	and the blind spot are color-coded as in Fig.~\ref{fig:m1mu}, with the 
	cross-hatched region in dark-blue corresponding to CMS limits~\cite{CMS:2013hja} on 
	$H,A\to \tau^+\tau^-$.  The region to the left of the dark-red dashed line at 
	$m_A \simeq m_{H^+} \simeq 480$ GeV is excluded by $B\to X_s\gamma$~\cite{Misiak:2015xwa}.}
	\label{fig:mAtb}
\end{figure}

Let us now examine the experimental limits associated with $\chi$--xenon scattering.  
In Fig.~\ref{fig:mAtb}, we show constraints in the $(m_A,\tan\beta)$ plane for two 
benchmark values of $M_1$ and $\mu$.  We find that as the mass splitting between 
$M_1$ and $\mu$ is decreased, the limits become significantly stronger.  This is 
because the amplitude $C_{q_i}^{h,H}$ scales like~$\sim 1/(M_1^2-\mu^2)$, so the 
naturalness requirement of light Higgsinos implies strong constraints on the SI 
cross section.  We also find blind spots similar to those previously 
identified~\cite{Ellis:2000ds,Ellis:2000jd,Baer:2006te,Huang:2014xua,Anandakrishnan:2014fia}, 
and see that the strongest limits are due to $H,A\to\tau^+\tau^-$ 
searches~\cite{CMS:2013hja} as one approaches (\ref{eq:h H blind analytic}) from 
below.

\begin{figure}[t]
	\centering\includegraphics[scale=0.9]{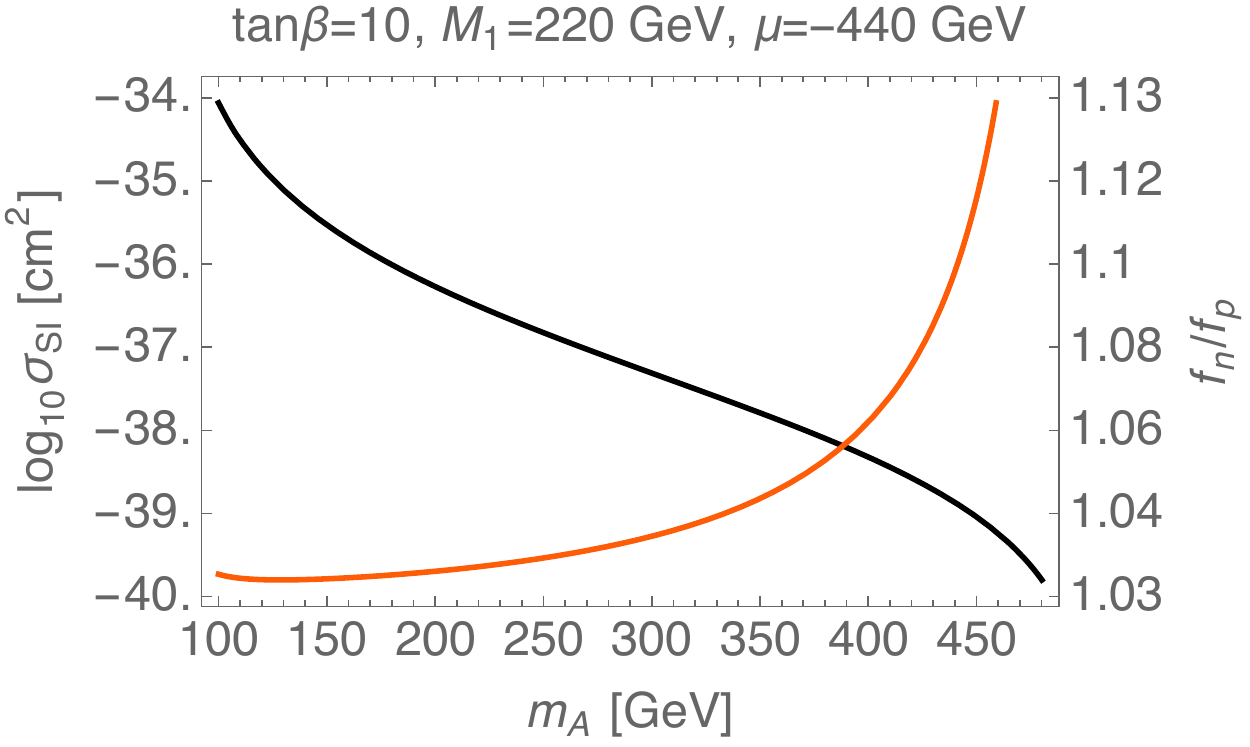}
	\caption{Pseudoscalar Higgs mass $m_A$ dependence of the SI $\chi$--xenon cross section  
	(black) and the central value of $f_n/f_p$ (red) as determined by~\sone.}
	\label{fig:h H cross_IV}
\end{figure}

What about isospin violation in this model?  Unlike single $h$ exchange, where $f_n/f_p$ 
is entirely fixed (\ref{eq:h ratio}) by hadronic quantities, the blind spot 
(\ref{eq:h H blind analytic}) for light and heavy Higgs bosons involves destructive 
interference between the respective amplitudes.  In general, we find that isospin 
violation can be enhanced in the vicinity of such blind spots because $f_n/f_p$ becomes 
sensitive to the scalar matrix elements and their uncertainties (\ref{eq:fNofP uncertainty}).  
This is evident in Fig.\ \ref{fig:h H cross_IV}, where the \textit{central value} of $f_n/f_p$ 
(determined by~\sone) reaches $\approx 15\%$ as the blind spot is approached with increasing $m_A$.

\begin{figure}[t]
	\centering
	\includegraphics[scale=0.9]{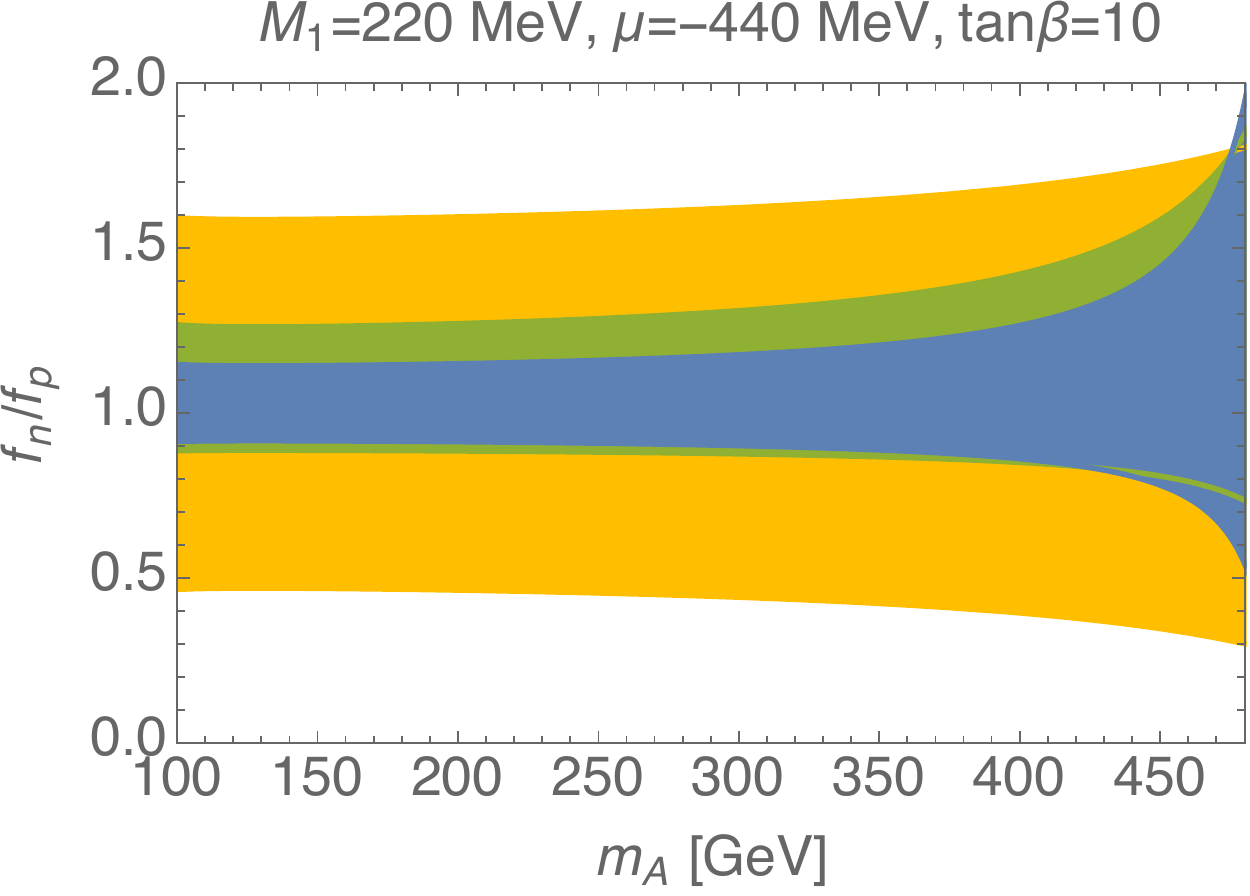}
	\caption{Amount of isospin violation in terms of $f_n/f_p$ due to $h,H$ exchange 
	in $\chi$--xenon scattering.  The shaded regions show the uncertainty on 
	$f_n/f_p$ due to each determination of the scalar matrix elements listed in 
	Sec.~\ref{sec:hadronic}.  Color coding as in Fig.\ \ref{fig:hNohP}.}
	\label{fig:HhIV_compare}
\end{figure}

In Fig.~\ref{fig:HhIV_compare} we compare the amount of isospin violation allowed by 
Methods 1-3. As observed in single $h$ exchange, the uncertainties associated with \stwo~are large, and differ by a factor of two or more for $m_A \lesssim 400$ GeV.  For $m_A \gtrsim 400$ GeV, a comparison between the Methods is obscured by the fact that the location of the blind spot is shifted depending on deviations from $D_N=U_N$ (Fig.\ \ref{fig:DNoUN}).

\begin{figure}[t]
	\centering
	\subfloat{\includegraphics[scale=0.6]{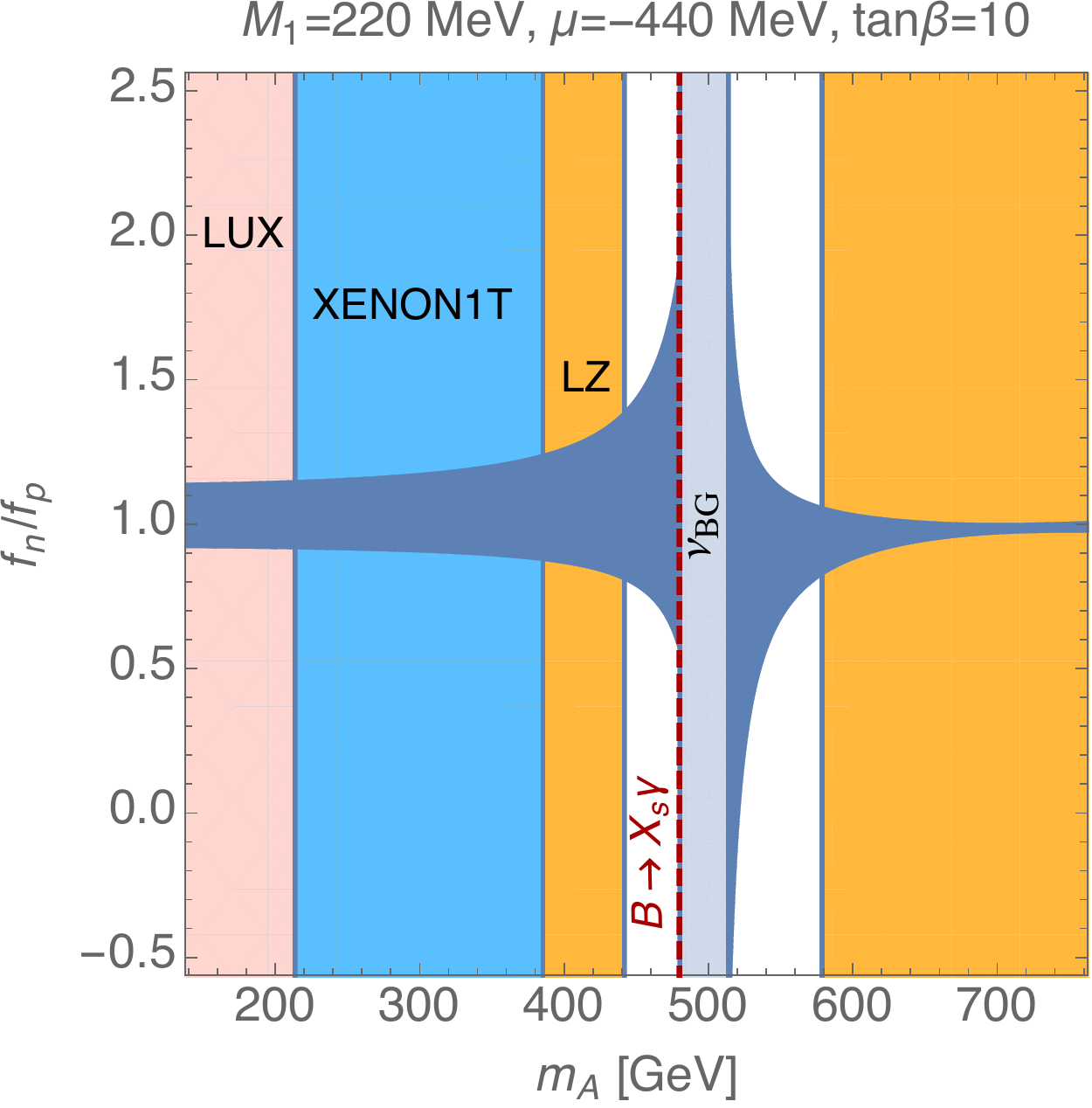}}
	\quad
	\subfloat{\includegraphics[scale=0.6]{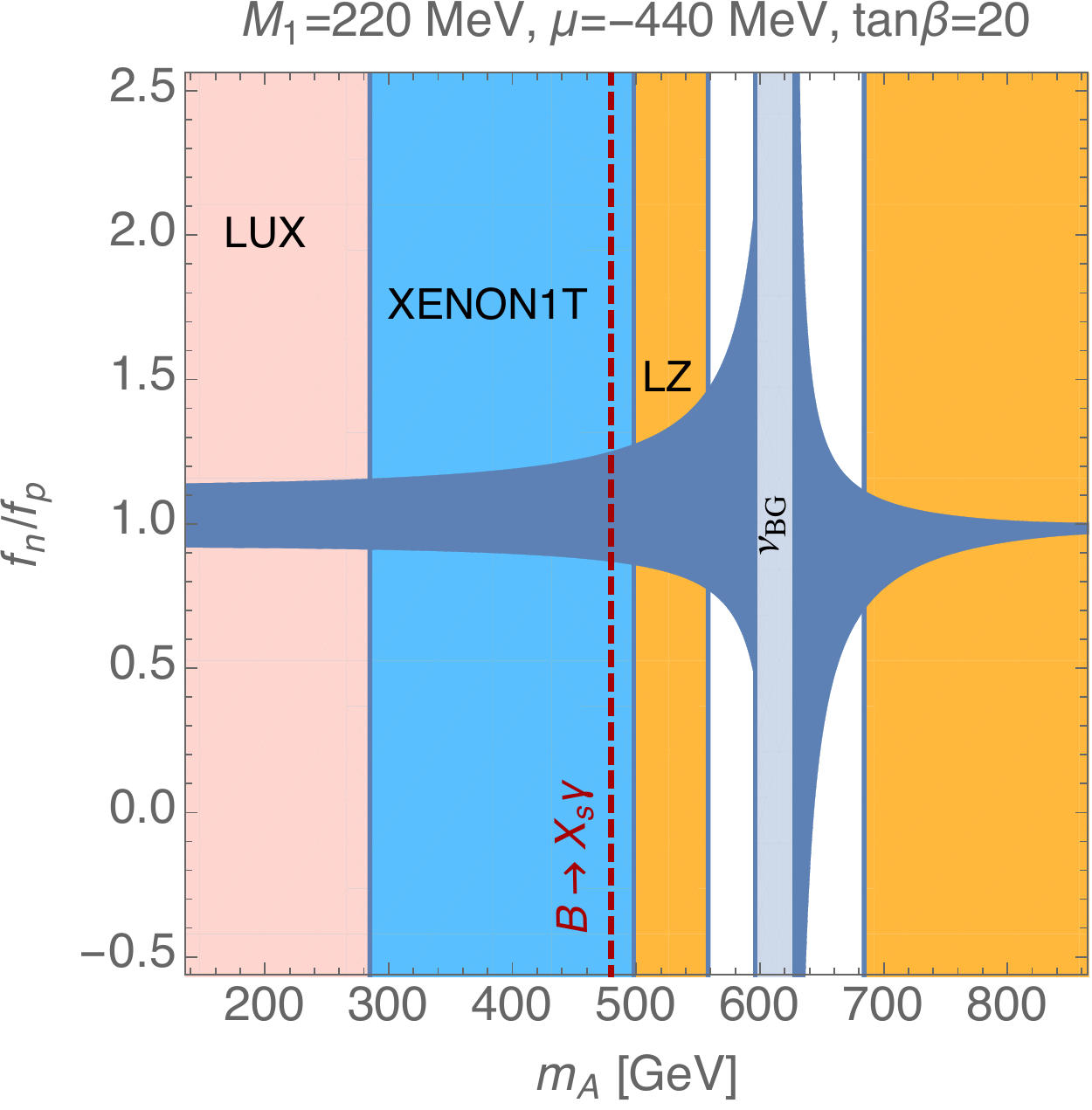}}
	\caption{Current and projected results for the measure $f_n/f_p$ of isospin 
	violation arising from $h,H$ exchange in SI $\chi$--xenon scattering for 
	$\tan\beta=10$ (left) and 20 (right).  The dark blue region denotes the 
	uncertainty on $f_n/f_p$ as determined by \sone.  The existing and projected 
	experimental limits are color coded as in Fig.\ \ref{fig:m1mu}, with the irreducible neutrino 
	background ($\nu$BG) shown by the central gray band. The region to the left of the dark 
	red dashed line at $m_A \simeq 480$ GeV is excluded by $B\to X_s\gamma$~\cite{Misiak:2015xwa}.}
	\label{fig:HhIv}
\end{figure}

In Fig.\ \ref{fig:HhIv} we display the allowed ranges of isospin violation due to \sone~for two values of $\tan\beta$.  For $\tan\beta=10$, the recent limit from 
$B\to X_s\gamma$~\cite{Misiak:2015xwa}  
excludes the region below the blind spot at $m_A \approx 500$ GeV.  Above 
the blind spot, the absence of a signal at LZ would imply that 
isospin violation as large as  
10\% becomes allowed.  For $\tan\beta=20$, the current limit from 
LUX~\cite{Akerib:2013tjd} allows around 10\% isospin violation, 
 although this occurs at a value of $m_A \approx 200$ GeV already excluded by the limits from 
$H,A\to\tau^+\tau^-$~\cite{CMS:2013hja} and $B\to X_s\gamma$~\cite{Misiak:2015xwa}.  The absence of a signal at XENON1T would 
allow  $\approx 20\%$, while at LZ this would imply that
isospin violation as large as $40\%$ is allowed within the uncertainties as one approaches the blind spot from below. 
As illustrated in Fig.~\ref{fig:h H cross_IV}, it is important to note 
that not only the allowed range,
but also the central value of $f_n/f_p$ can increase as the blind spot is approached. 
In consequence, the absence of signals in SI DM searches pushes the parameter space 
into blind spots, at which $f_n/f_p$ may become large and thus the accurate determination 
of $\delta f_n$ and $\delta f_p$ becomes paramount.  A comparison for other nuclear targets can be inferred by taking the limits on $f_n/f_p$ and comparing against Fig.\ \ref{fig:target IV}.

\subsection{SM-like Higgs and light squark exchange}
In the previous subsection, we investigated parameter configurations~\eqref{eq:h H blind} where the $h,H$ amplitudes  
interfere destructively and observed that isospin violation can be enhanced in the proximity of these blind spots
(Fig.\ \ref{fig:HhIv}).  Next, we examine if blind spots still exist once third-generation squarks are added to the spectrum of model (A).

The effects on the SI amplitude due to squarks from the first two generations 
were considered in~\cite{Cheung:2012qy} (including $h$ exchange) and shown to be small 
due to the stringent limits from LHC searches.  However, the existing limits on 
third-generation squark masses are much weaker, so that effects from stops and 
sbottoms can be significantly larger.

The simplest model, i.e.\ with minimal particle content, would involve a single, mostly right-handed stop $\tilde{t}_R$.   
However, as one can see from~\eqref{up-squark-contribution}, this contribution is 
not $\tan\beta$ enhanced and thus the $h$ contribution (\ref{eq:rh}) dominates 
the SI cross section.  Therefore, we consider a spectrum where $h$ and 
$\tilde{t}_{1,2}$ are the dynamical degrees of freedom [model (C) in Fig.\ 
\ref{fig:simple}]. Since a left-handed stop is always associated with a left-handed 
sbottom $\tilde{b}_L$, the sbottom contribution must be included as well. Although 
this does not increase the number of free parameters, we can see from  
(\ref{down-squark-contribution}) that the sbottom amplitude is $\tan\beta$ 
enhanced and, crucially, \emph{can} compete with the Higgs contribution to the SI 
amplitude.  Note that while the Higgs amplitude vanishes with decoupling 
Higgsinos (thereby violating the minimal naturalness conditions), this is not the 
case for the sbottom contribution, which possesses a term proportional to $\mu$.

Using our simplified expressions, we find that a blind spot occurs if the condition 
\begin{equation}
	\frac{6}{m_h^2}(M_1+\mu s_{2\beta}) \bigg[ \sum_q f_q^N + 3f_Q^N \bigg] 
	- \frac{1}{M_1^2-m_{\tilde{b}_L}^2} (M_1+\mu\tan\beta) f_Q^N \simeq 0
	\label{eq:h squark blind}
\end{equation}
is satisfied. Here, we have ignored the numerically small stop contribution 
$C_t^{\tilde{t}}$ and approximated the effects due to sbottom loops (\ref{K function}) 
by the tree-level propagator. This latter approximation is for illustrative 
purposes only, and in our numerical analysis we use the exact one-loop expressions.  
Using the scalar matrix elements from \sone, we have 
$\sum_q f_q^N + 3f_Q^N \simeq \tfrac{1}{3}$ and $f_Q^N \simeq \tfrac{1}{15}$. Therefore, 
for moderate to large values of $\tan\beta$ the blind spot condition simplifies to
\begin{equation}
	\frac{30}{m_h^2} (M_1 + \mu s_{2\beta}) 
	+ \mu \tan\beta \frac{1}{m_{\tilde{b}_L}^2-M_1^2} \simeq 0\,.
	\label{eq:h squark blind analytic}
\end{equation}
As expected, this blind spot shares common features with the one found~\cite{Huang:2014xua} for $h,H$ exchange (\ref{eq:h H blind analytic}): it requires negative values of $\mu$, so that the couplings to $h$ are suppressed and destructive interference between the $h$ and $\tilde{b}_L$ amplitudes can occur. However, larger values of $|\mu|$ are required in order to overcome 
the factor of 30 in the Higgs amplitude. 

Before determining the experimental limits on this model, let us consider the size of the  parameter space. The $h$ amplitude depends on $M_1,\,\mu,$ and $\tan\beta$, so we need to add the 
parameters $(\textbf{m}_Q)_{33},(\textbf{m}_U)_{33}$, and $X_t$ of the stop mass matrix.  As noted above, the sbottom contribution does not involve additional 
parameters since the left-handed sbottom mass is given by
\begin{equation}
	m_{\tilde{b}_L}^2 = ({\bf m}_Q)_{33}^2 + m_b^2 - (\tfrac{1}{2} - \tfrac{1}{3}s_W^2)m_Z^2 c_{2\beta}\,.
\end{equation}
To reduce the number of free parameters we fix $(\textbf{m}_Q)_{33} \approx (\textbf{m}_U)_{33}$, in which case the left- and right-handed entries in the stop mass matrix become nearly degenerate, while 
the physical mass eigenvalues read
\begin{equation}
	m_{\tilde{t}_1,\tilde{t}_2}^2 = m_{\tilde{t}_L}^2 \mp m_tX_t\,.
\end{equation}
This allows us to express our simplified expressions (\ref{up-squark-contribution}-\ref{down-squark-contribution}) in terms of the physical masses and compare with collider 
limits in the $(m_{\tilde{t}_1},M_1)$ plane. In order for light stops to generate the correct Higgs mass in the MSSM, we assume these states are mixed in such a way so as to give a maximal contribution to the Higgs mass. This is achieved by noting that the one-loop stop contribution to the Higgs mass 
\begin{equation}
	m_h^2 \approx m_Z^2 c_{2\beta}^2 + \frac{3}{4\pi^2}\frac{\bar{m}_t^4}{v^2} 
	\left[ \ln \frac{\hat{m}_{\tilde{t}}^2}{\bar{m}_t^2} + \frac{X_t^2}{\hat{m}_{\tilde{t}}^2} 
	\left(1-\frac{X_t^2}{12\hat{m}_{\tilde{t}}^2} \right) \right]
\end{equation}
is maximized at $|X_t|=X_t^\mathrm{max}=\sqrt{6}\hat{m}_{\tilde{t}}$, where 
$\hat{m}_{\tilde{t}}^2=m_{\tilde{t}_1}m_{\tilde{t}_2}$ is the average stop mass. Since $m_h$ is bounded at tree-level by $m_Z$, requiring $m_h \simeq 125 $ GeV implies a lower bound $\hat{m}_{\tilde{t}}^2 \gtrsim 550$~GeV for $X_t^\mathrm{max}$. While this is the main source of fine-tuning in the MSSM, it can be easily evaded in e.g.\ non-minimal SUSY models, where the correct Higgs mass can be obtained via non-decoupling $D$-terms~\cite{Batra:2003nj}, or in the next-to-minimal supersymmetric SM with special parameter choices~\cite{Hall:2011aa}.

\begin{figure}[t]
	\centering\includegraphics[scale=1]{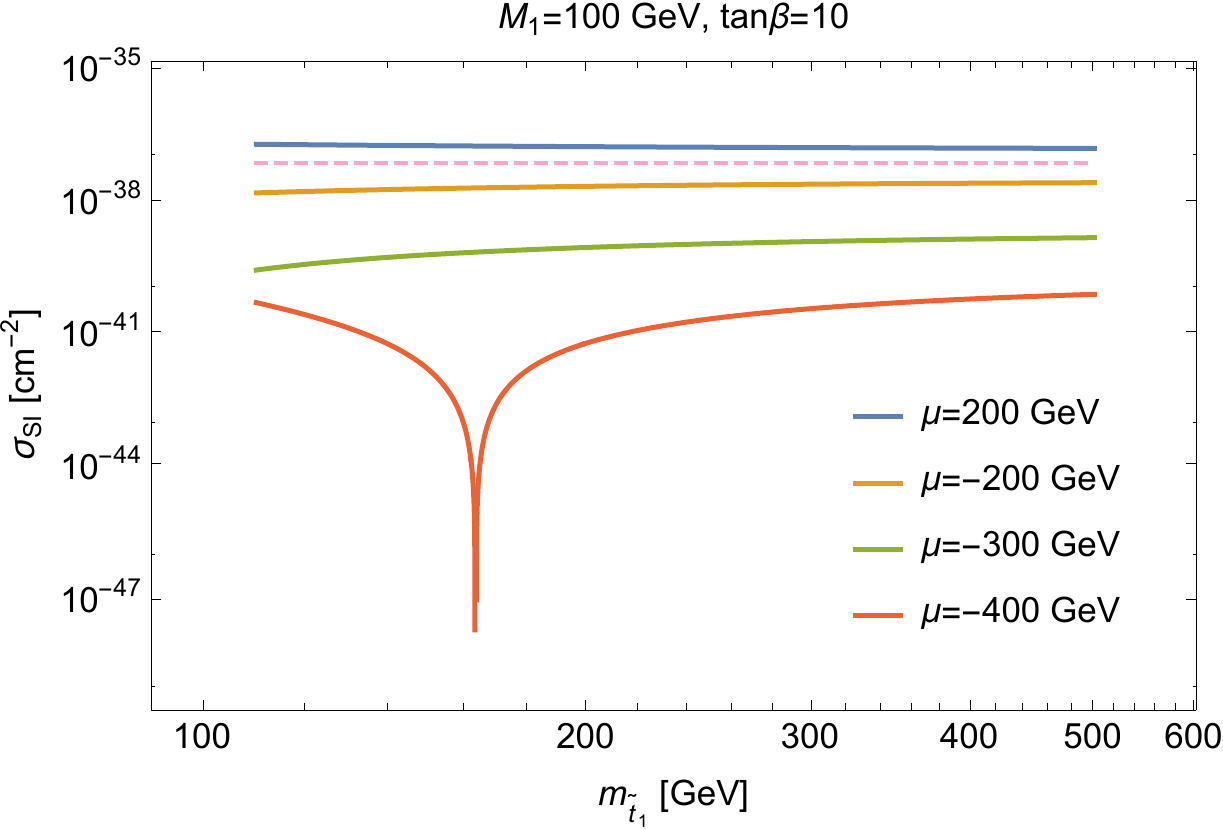}
	\caption{SI $\chi$--xenon cross sections as a function of the lightest stop mass 
	$m_{\tilde{t}_1}$.  The solid lines correspond to different values of the Higgsino 
	mass $\mu$, while the pink dashed line is the existing limit from LUX.  The 
	blind spot at $m_{\tilde{t}_1} \simeq 160$ GeV is shown in red.}
	\label{fig:stop_cross}
\end{figure}

In Fig.\ \ref{fig:stop_cross} the SI cross section is displayed as a function 
of the lightest stop mass $m_{\tilde{t}_1}$ for $\tan\beta=10$ and several values 
of $\mu$.  We find that a positive value of $\mu$ is excluded by LUX, while negative values become progressively harder to constrain as the mass difference between the bino and Higgsinos is increased.  The blind spot is clearly seen for $\mu=-4M_1$ and occurs at a light stop mass $m_{\tilde{t}_1}\simeq 160$ GeV.

\begin{figure}[t]
	\centering
        \subfloat{\includegraphics[scale=0.8]{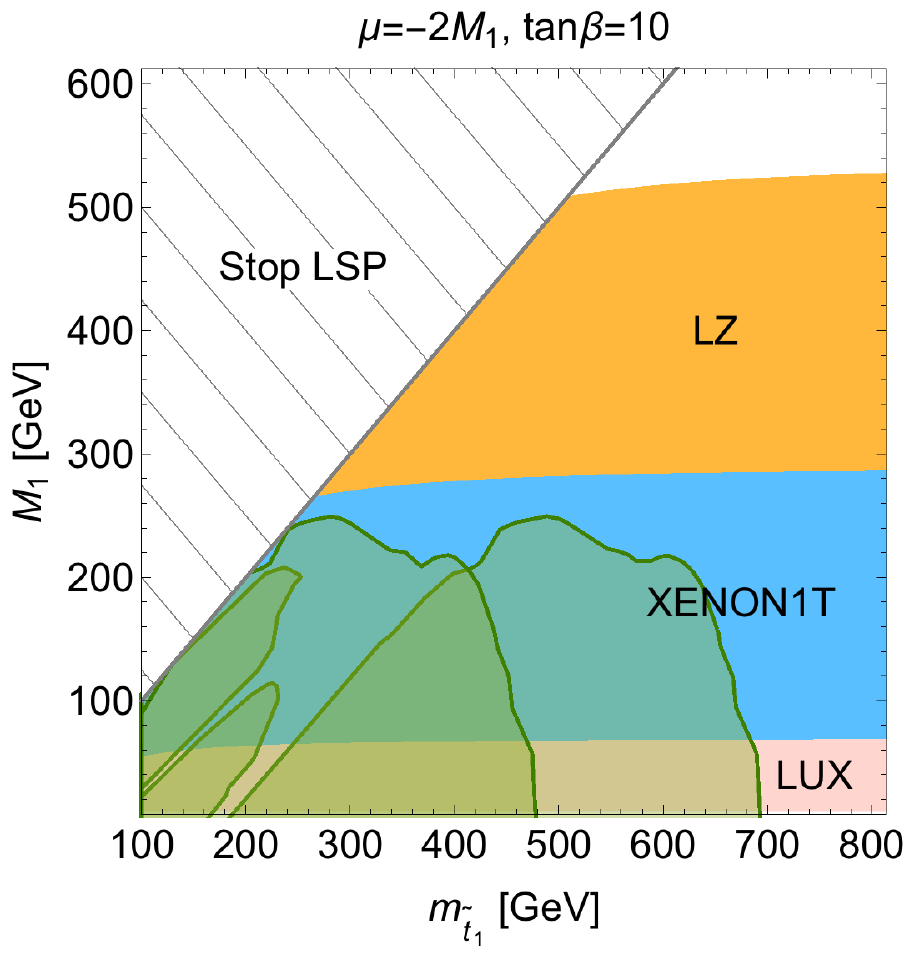}}%
        \quad
        \subfloat{ \includegraphics[scale=0.8]{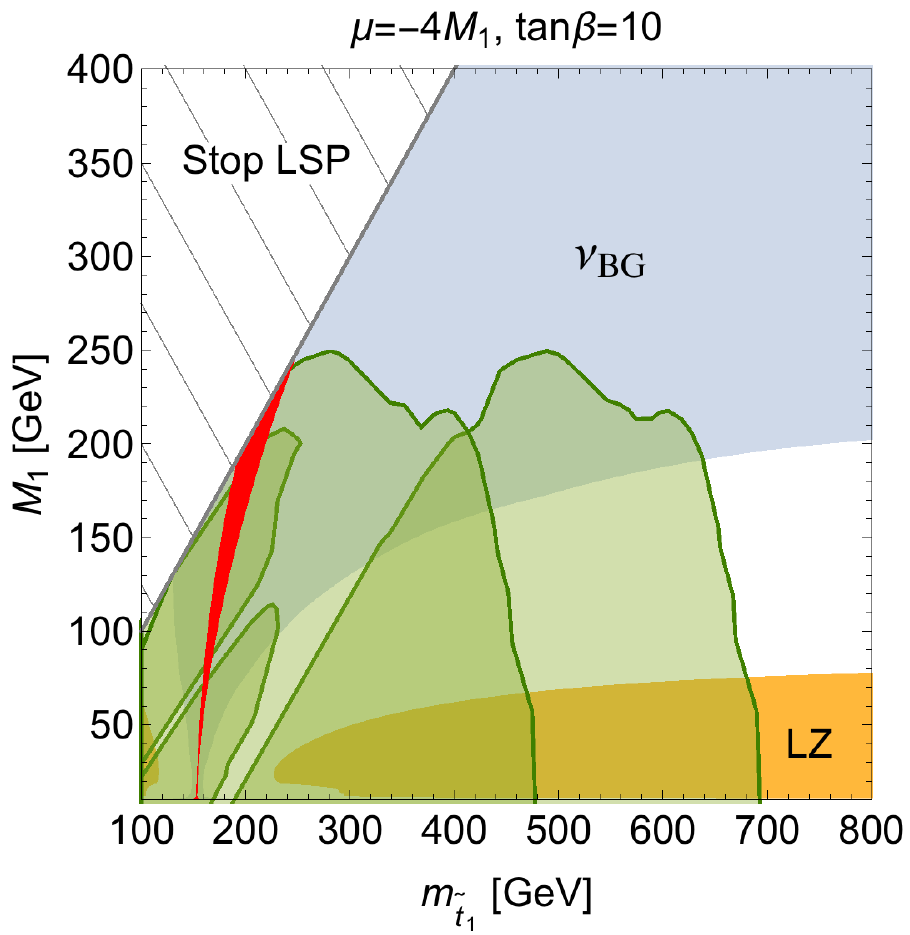}}
        \quad
        \subfloat{\includegraphics[scale=0.8]{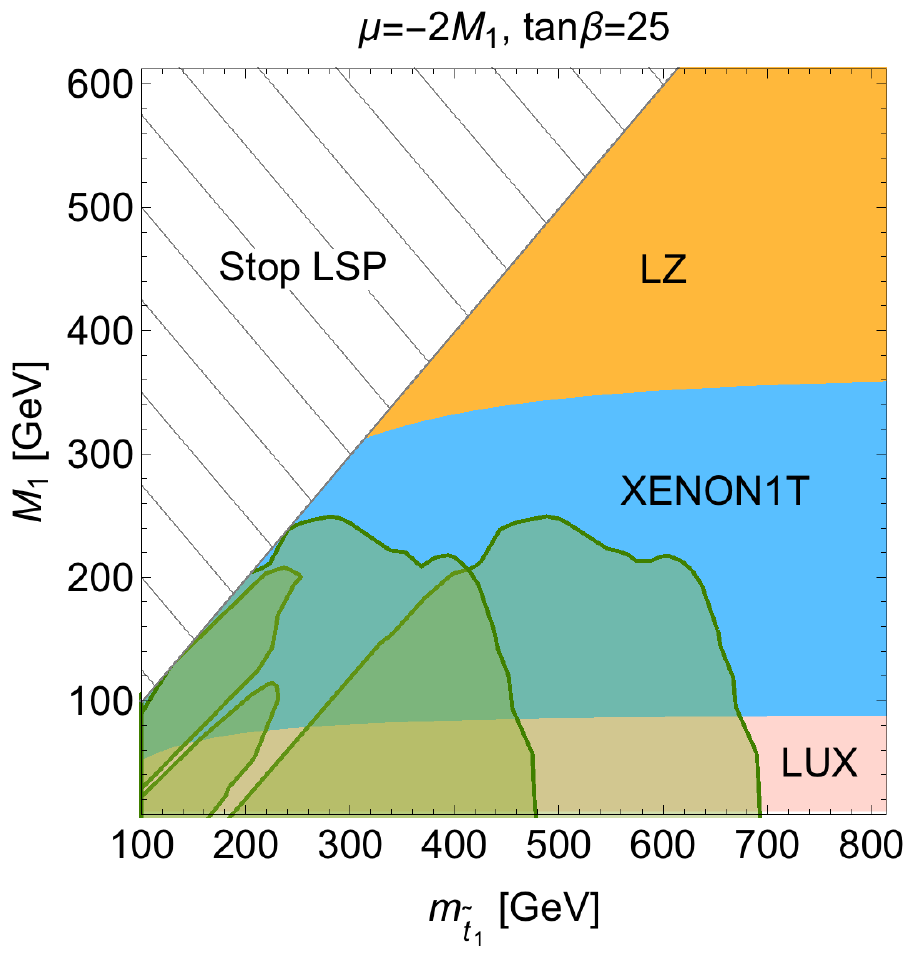}}%
        \quad
        \subfloat{\includegraphics[scale=0.8]{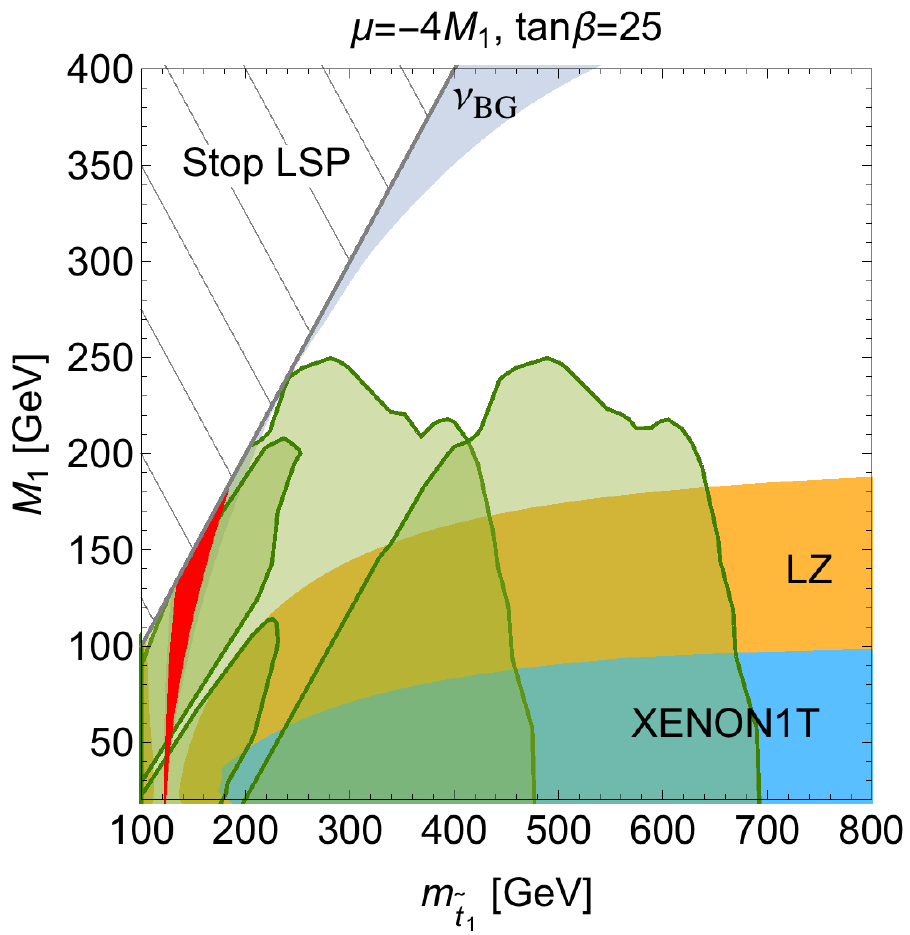}}
	\caption{Current and projected limits due to light Higgs $h$ and squark 
	$\tilde{t}_{1,2},\tilde{b}_L$ exchange in $\chi$--xenon scattering with 
	$\tan\beta=10$ (top row) and $\tan\beta=25$ (bottom row) and two benchmarks for negative $\mu$. The figures show the DM 
	constraints from LUX~\cite{Akerib:2013tjd} and XENON1T~\cite{Aprile:2012zx} (color 
  coded as in Fig.~\ref{fig:m1mu}), while limits from direct searches for stops 
  and sbottoms at ATLAS~\cite{ATLAS:2013pla} are shown in green.  
  The blind spot is shown in red and lies within the neutrino background ($\nu_\mathrm{BG}$) shown in gray. The hatched region corresponds to the case where $\tilde{t}_1$ becomes the LSP.}
	\label{fig:M1mt1tb10}
\end{figure}

In Fig.\ \ref{fig:M1mt1tb10} we show the interplay between collider and DM direct-detection limits in the $(m_{\tilde{t}_1},M_1)$ plane for two values of $\tan\beta$ and $\mu$.  We find that in the absence of blind spots $(\mu=-2M_1)$, LUX excludes the $M_1 \lesssim 50$ GeV region across a large range of stop masses.  These limits will be significantly improved if XENON1T and LZ do not detect a DM signal, with whole regions below $M_1 \approx 300$ GeV and 500 GeV excluded by the respective experiments.  In these cases, the direct-detection limits surpass those derived from the ATLAS searches for stops and sbottoms. 

In the blind spot region $(\mu=-4M_1)$, the DM limits are considerably weakened, with LZ excluding most of the $M_1\lesssim 50$ MeV and 150 MeV regions for $\tan\beta=10$ and 25 respectively. As noted in Sec.\ \ref{sec:h H exchange}, isospin violation can be enhanced in the proximity of blind spots which arise from destructive interference in the amplitude.  Since this is a generic feature of SI scattering, it follows that isospin violation can also be large near the blind spots shown in Fig.\ \ref{fig:M1mt1tb10}.  Although these regions are excluded by collider limits, we have not ruled out the possibility that blind spots for this model occur in viable regions of parameter space.  

\subsection{Generic Higgs and light squark exchange}
In this Section, we consider the effect of adding $H$ to the particle content, so that the active degrees of freedom are $h,H$ and $\tilde{t}_{1,2},\tilde{b}_L$ [model (D) in Fig.\ \ref{fig:simple}]. This is the most ``natural'' model studied in this article since a large value of $m_H$ would also require fine tuning~\cite{Katz:2014mba}.

To derive an analytic formula for the blind spot, we follow the same steps used to obtain (\ref{eq:h squark blind}).  For moderate to large values of $m_A>m_h$ and $\tan\beta$, we find that a blind spot occurs whenever
\begin{equation}
	\frac{30}{m_h^2} (M_1 + \mu s_{2\beta}) 
	+ \mu\tan\beta \left(\frac{15}{m_H^2} +\frac{1}{m_{\tilde{b}_L}^2-M_1^2}\right) \simeq 0
	\label{eq:hH squark blind}
\end{equation}
is satisfied.  In this case, the inclusion of $H$ has the effect of shifting the location of the blind spot found for $h$ and squark exchange (\ref{eq:h squark blind analytic}).  In particular, negative values of $\mu$ are still required.
\begin{figure}[t]
	\centering  
        \subfloat{\includegraphics[scale=0.75]{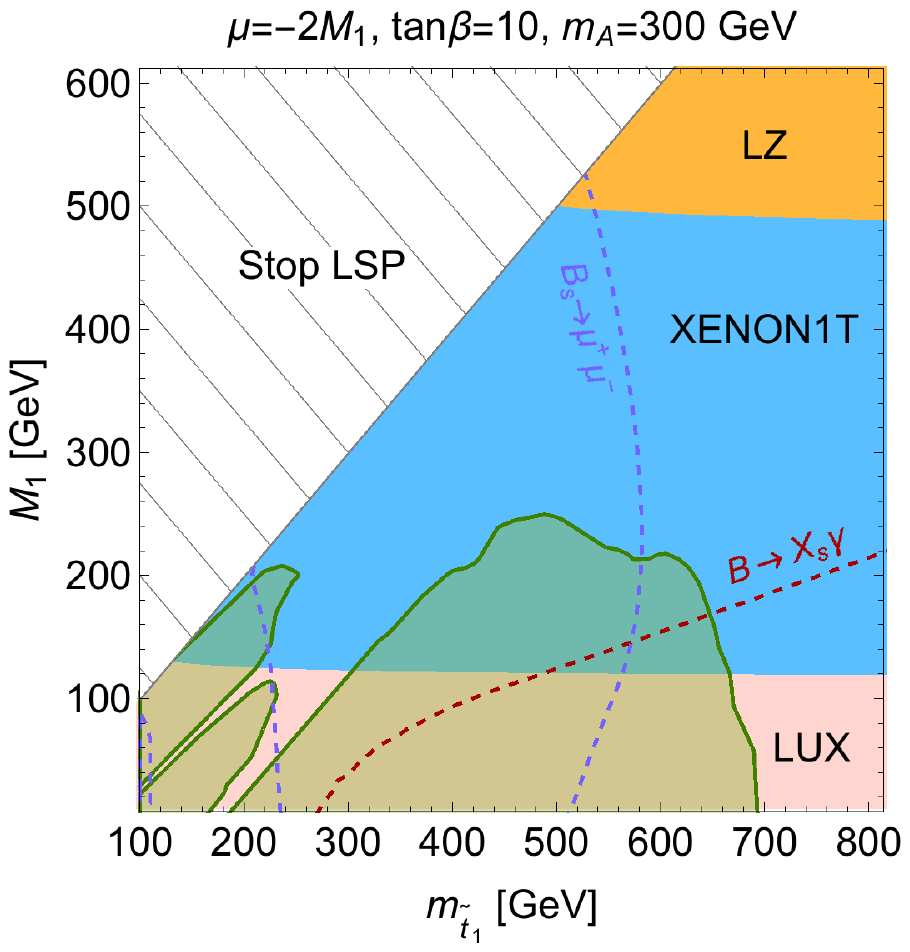}}
        \quad
        \subfloat{\includegraphics[scale=0.75]{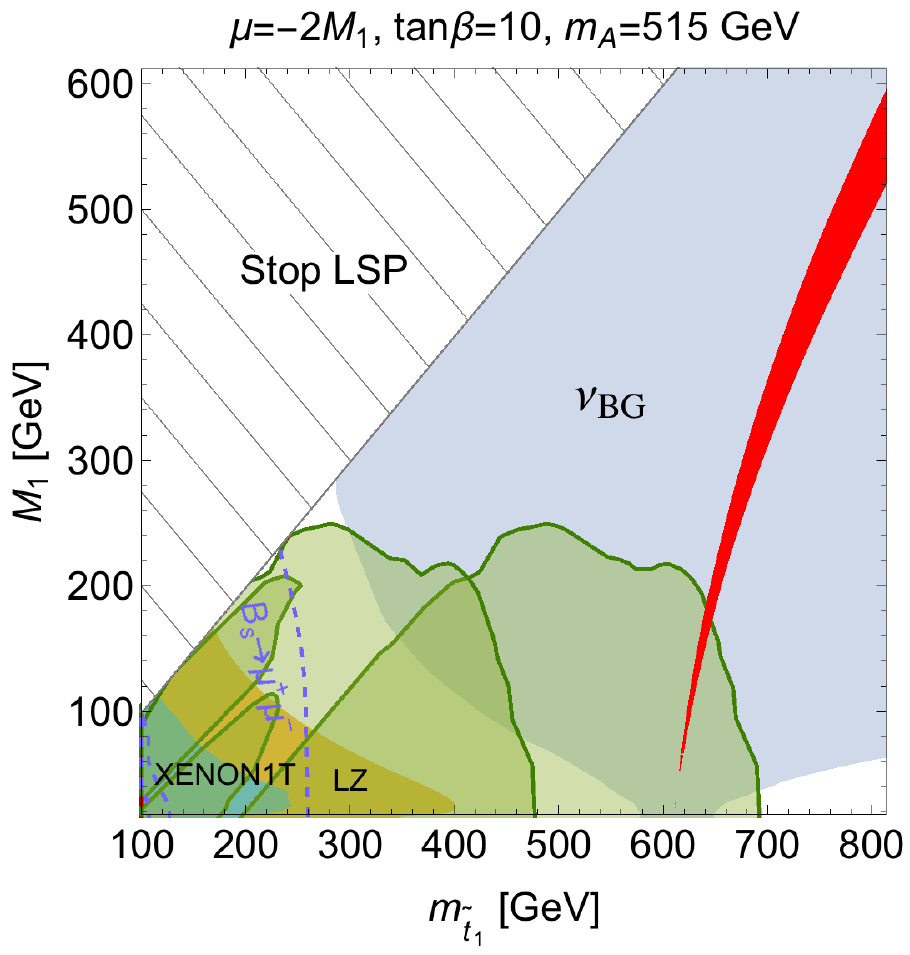}}
        \quad
        \subfloat{\includegraphics[scale=0.75]{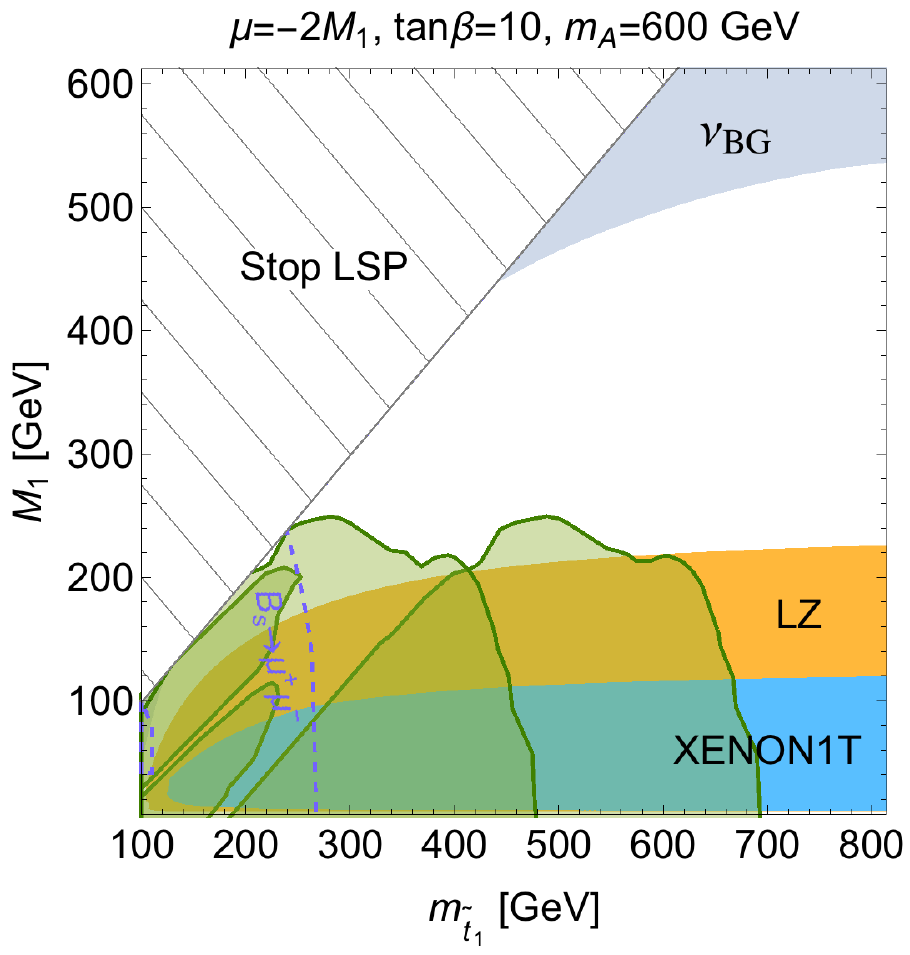}}
	\caption{Current and projected limits in the $(m_{\tilde{t}_1},M_1)$ plane from 
	$h,H$ and $\tilde{t}_{1,2},\tilde{b}_L$ exchange in $\chi$--xenon scattering.  
	The three plots represent different values of $m_A$, for fixed $\tan\beta=10$.  For $A_t^\mathrm{max}<0$, the regions between the purple dashed lines are excluded by $B_s\to\mu^+\mu^-$, while regions to the left of the dark red  dashed  lines are excluded by $B\to X_s\gamma$.  
	Excluded regions from direct detection are color-coded as in Fig.\ \ref{fig:M1mt1tb10}.}
	\label{fig:M1mt1mAtb10}
\end{figure}

To examine the limits on this model, we must also consider flavor observables since $H$ and light stops contribute to $B\to X_s\gamma$ and $B_s\to\mu^+\mu^-$. To evaluate the flavor constraints we use \textsc{susy\_\,flavor-2.51}~\cite{Rosiek:2010ug,Crivellin:2012jv,Rosiek:2014sia} and implement the NNLO SM calculation by constraining the ratio
\begin{equation}
R_{\rm SUSY}(B\to X_s\gamma)=\dfrac{{\rm Br}(B\to X_s\gamma)^{\rm MSSM}_{\rm SUSY\underline{\;\;}FLAVOR}}{{\rm Br}(B\to X_s\gamma)^{\rm SM}_{\rm SUSY\underline{\;\;}FLAVOR}}
\end{equation}
to lie within the allowed range for 
\begin{equation}
R_{\rm EXP}(B\to X_s\gamma)=\dfrac{{\rm Br}(B\to X_s\gamma)_{\rm EXP}}{{\rm Br}(B\to X_s\gamma)^{\rm SM}_{\rm NNLO}}\,.
\end{equation}
Here we use the recent calculation of~\cite{Misiak:2015xwa}
\begin{equation}
	\mathrm{Br}(B\to X_s\gamma)_{\mathrm{SM}} = (3.36 \pm 0.23) \times 10^{-4} 
\end{equation}
to incorporate the NNLO SM prediction. For the experimental value we use the PDG average~\cite{PDG} within 2$\sigma$ uncertainties,
\begin{equation}
	\mathrm{Br}(B\to X_s\gamma)_{\rm EXP} = (3.41 \pm 0.21 \pm 0.07) \times 10^{-4}\,.
\end{equation}
We add the theoretical error linearly with the experimental one, so that $R_\mathrm{SUSY}$ is required to lie within the interval
\begin{equation}
	R_{\rm EXP}^{\rm MIN}(B\to X_s\gamma)\leq R_{\rm SUSY}(B\to X_s\gamma)\leq R_{\rm EXP}^{\rm MAX}(B\to X_s\gamma)\,.
\end{equation}
For $B_s\to\mu^+\mu^-$ we adopt the same procedure to impose limits on the relevant SUSY parameter space. Here the 
SM prediction~\cite{Bobeth:2013uxa} is 
	\begin{equation}
		\mathrm{Br}(B_s \to \mu^+\mu^-)_\mathrm{SM} = (3.65 \pm 0.23) \times 10^{-9}\,,
	\end{equation}
and has to be compared against	
	\begin{equation}
		\mathrm{Br}(B_s \to \mu^+\mu^-)_{\rm EXP} = (3.1 \pm 0.7) \times 10^{-9}\,.
	\end{equation}

In Fig.~\ref{fig:M1mt1mAtb10} we display limits in the $(m_{\tilde{t}_1},M_1)$ plane for $\tan\beta=10$ and several values of $m_A$.  Since this choice of $\tan\beta$ corresponds to a horizontal slice through the $h,H$ parameter space (Fig.\ \ref{fig:mAtb}), the effect of increasing $m_A$ is to probe the effect of the $h,H$ blind spot (\ref{eq:h H blind analytic}) from below.  For $m_A=300$ GeV and away from the blind spot, we find that LUX excludes the band below $M_1 \approx 100$ GeV.  In this case, the limits from $B\to X_s\gamma$ and $B_s\to\mu^+\mu^-$ provide a complementary and stringent constraint: compatibility with both observables and LUX only leaves a small region of parameter space viable.  XENON1T and LZ will carve out most of the remaining parameter space, providing a very strong constraint on the light $m_A$ scenario.  However, for heavier values of $m_A$, the constraints from flavor and direct detection weaken considerably and here the collider bounds become the dominant constraint.  This is particularly evident in the second plot of Fig.\ \ref{fig:M1mt1mAtb10}, where the blind spot suppresses the SI cross section and the moderate value of $m_A$ reduces the tension with $B\to X_s\gamma$ entirely.

\begin{figure}[t]
	\centering
        \subfloat{\includegraphics[scale=0.8]{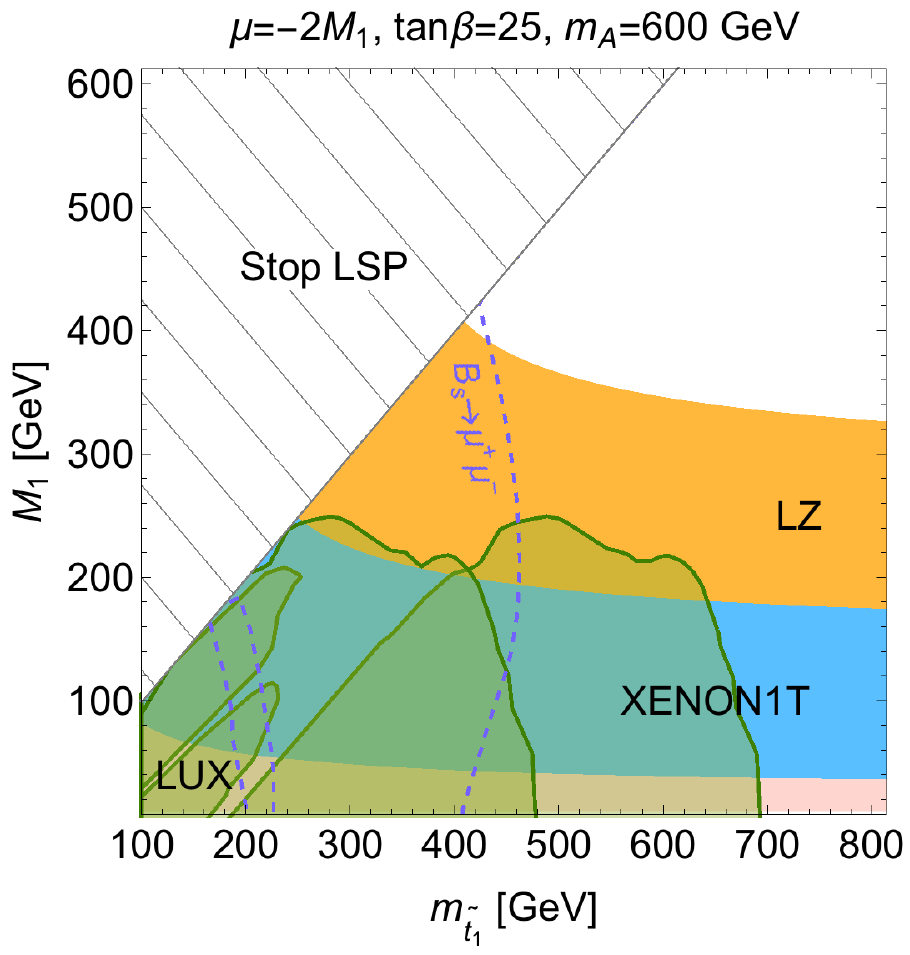}}
        \quad
        \subfloat{\includegraphics[scale=0.8]{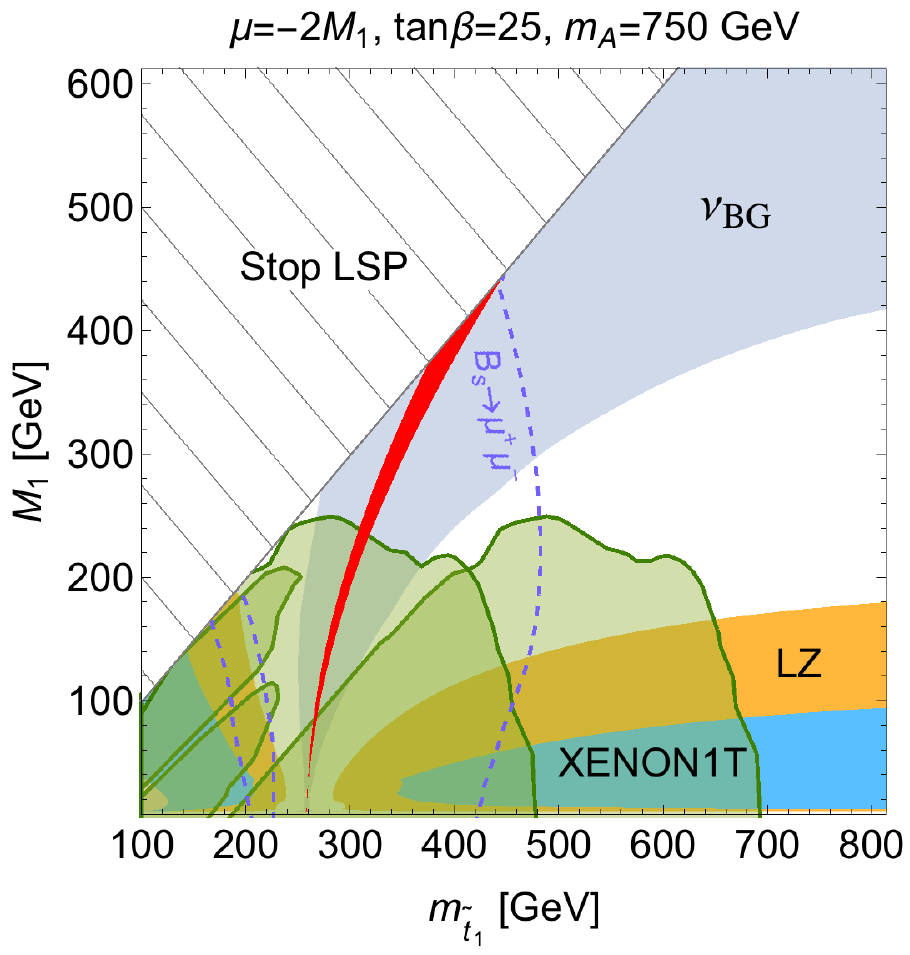}}
	\caption{Current and projected limits in the $(m_{\tilde{t}_1},M_1)$ plane from 
	$h,H$ and $\tilde{t}_{1,2},\tilde{b}_L$ exchange in $\chi$--xenon scattering.  
	In the figures, the value of $m_A$ is increased for fixed $\tan\beta$.  
	Excluded regions are color-coded as in Fig.\ \ref{fig:M1mt1mAtb10}, with $B\to X_s\gamma$ ruling out both values of $m_A$ entirely.}
	\label{fig:M1mt1mAtb25}
\end{figure}

We consider the effect of increasing $\tan\beta$ in Fig.\ \ref{fig:M1mt1mAtb25}.  For $m_A=600$ GeV, there is no blind spot in the physical region, with most of the area below $m_{\tilde{t}_1} \approx 450$ MeV excluded, while for $m_A=750$ GeV a blind spot does occur.  This latter region is allowed by current collider limits, but excluded by $B\to X_s\gamma$.  However, one should keep in mind that the flavor bounds included in Fig.~\ref{fig:M1mt1mAtb10} (and Fig.\ \ref{fig:M1mt1mAtb25}) are the least rigorous ones as they can 
be evaded if some of the underlying assumptions are relaxed: in the presence of non-minimal sources of flavor violation the bounds can become weaker. In fact, a mass splitting among the left-handed squarks deviates from naive minimal flavor violation since there are either off-diagonal elements in the up or in the down sector of the squark mass matrix (or in both simultaneously). Furthermore, relaxing our assumption that the left-handed bilinear terms are diagonal in the down basis~(\ref{eq:sqrk mass}) would lead to additional effects in flavor observables.

\section{Conclusions}
\label{sec:conclusions}
In this paper we examined four simplified models in the framework of the MSSM 
where all but a handful of superpartners are decoupled from the spectrum. We started from the minimal model necessary to provide a viable DM candidate and sequentially added particles to render the spectrum more natural. The key result, analytic expressions for the Wilson coefficients relevant for the Higgs- and squark-exchange contribution to spin-independent WIMP--nucleon scattering, is summarized in~(\ref{up-squark-contribution}-\ref{eq:C Higgs dwn}).

As the main application of our scheme, we studied the amount of isospin violation generated by single-nucleon contributions to the spin-independent WIMP--nucleon cross section. In general, isospin violation is a rather small effect: for pure $h$ exchange, the amount of isospin violation is $\approx 5\%$, i.e.\ in line with common expectations and results in two-Higgs-doublet models~\cite{Drozd:2014yla}. Beyond single $h$ exchange, however, the effect can be enhanced in the proximity of blind spots. In such cases, the blind spots occur (at a given order in perturbation theory) due to destructive interference among different contributions to the SI amplitudes.  As a consequence, small variations of the amplitude become increasingly important. Although the SI cross sections are strongly suppressed in the vicinity of blind spots, as direct-detection experiments become increasingly more sensitive, MSSM models are pushed towards these corners of parameter space. For instance, we find that for $h$ and $H$ exchange, the projected limits from LZ~\cite{Malling:2011va} allow isospin violation to be as large as $40\%$, which increases rapidly as one approaches the irreducible neutrino background. In this way, precise predictions for isospin violation are essential to relate future direct-detection data to MSSM predictions. 

In our simplified models, the source of isospin violation originates purely from the SM; it is the blind spots that make these small differences prominent. This situation is unlike e.g.\ in $Z'$ models which introduce new sources of isospin violation beyond the SM.
Therefore, an accurate evaluation of isospin violation requires a careful assessment of the nuclear input quantities and the associated hadronic uncertainties. We demonstrated this point by comparing different methods to determine the proton and neutron scalar matrix elements that are currently used in the literature. While the traditional approach based on $\chi$PT$_3$ relations suffers from large and, in part, uncontrolled uncertainties, the hadronic input can be accurately evaluated by using the two-flavor formalism developed in~\cite{Crivellin:2013ipa}. In particular, we showed that in the three-flavor framework, depending on which input is used for the strangeness-related quantities, incorrect conclusions concerning both central values and uncertainties can occur.   

We also extended our models to include light stops and sbottoms in the spectrum. Again, for certain corners of the parameter space, cancellations occur that suppress the amplitude and led us to the identification of new blind spots. We identified these blind spots analytically in (\ref{eq:h squark blind analytic}) and (\ref{eq:hH squark blind}). 
Furthermore, the interplay between DM, collider, and flavor limits was studied, finding that the inclusion of the latter tends to exclude configurations with blind spots which are allowed by collider bounds.  Only for $\tan\beta=10$ and near the blind spot generated by $h$ and $H$ exchange, did we find a blind spot consistent with all constraints (Fig.\ \ref{fig:M1mt1mAtb10}). 

\section*{Acknowledgments}
We thank Ross Young and Joel Giedt for useful correspondence regarding~\cite{Giedt:2009mr}, Ahmed Ismail and Thomas Rizzo for providing us with details of their pMSSM benchmark models~\cite{Cahill-Rowley:2013gca}, Genevi\`eve B\'elanger and Alexander Pukhov for correspondence concerning \textsc{micrOMEGAs} \cite{Belanger:2013oya}, and Xavier Garcia i Tormo and DingYu Shao for useful discussions. 
Financial support by BMBF ARCHES, the Helmholtz Alliance HA216/EMMI, and the Swiss
National Science Foundation (SNF) is gratefully acknowledged. A.C.\ is supported by a Marie Curie Intra-European Fellowship of the European Community's 7th Framework Programme under contract number (PIEF-GA-2012-326948).

\appendix
\section{Perturbative diagonalization of the neutralino mass matrix}
In this Appendix, we diagonalize the neutralino mass matrix $M^\chi$ (\ref{eqn:mX}) 
via a perturbative expansion in $v/M_\mathrm{SUSY}$.  Following~\cite{Arnowitt:1995vg,Hofer:2009xb,Crivellin:2011jt}, we first consider the diagonal matrix $Z_a^{\chi\dagger} M^{\chi\dagger} M^\chi Z_a^\chi$, 
where to leading order in $v/M_\mathrm{SUSY}$ we have
\begin{equation}
	Z_a^\chi = \left( \begin{array}{cccc} 
	1 	& 	0 	& 	\frac{g_1v}{\sqrt{2}} \frac{(M_1^* c_\beta+ \mu s_\beta )}{|M_1|^2-|\mu|^2} 	
	&	-\frac{g_1v}{\sqrt{2}} \frac{(M_1^*s_\beta + \mu c_\beta)}{|M_1|^2-|\mu|^2} \\
	0 	& 	1 	&	-\frac{g_2v}{\sqrt{2}} \frac{(M_1^* c_\beta + \mu s_\beta )}{|M_1|^2-|\mu|^2} 	
	&	\frac{g_2v}{\sqrt{2}} \frac{(M_2^* s_\beta + \mu c_\beta)}{|M_1|^2-|\mu|^2} \\
	-\frac{g_1v}{\sqrt{2}} \frac{(M_1c_\beta + \mu^* s_\beta)}{|M_1|^2-|\mu|^2} 	
	& 	\frac{g_2v}{\sqrt{2}}\frac{(M_1 c_\beta+\mu^* s_\beta )}{|M_1|^2-|\mu|^2} 	
	& 	1 	& 	0 \\
	\frac{g_2v}{\sqrt{2}} \frac{(M_1s_\beta + \mu^* c_\beta)}{|M_1|^2-|\mu|^2} 
	&	 -\frac{g_2v}{\sqrt{2}} \frac{(M_2s_\beta + \mu^* c_\beta)}{|M_1|^2-|\mu|^2} 
	&	0 	& 	1 
		\end{array} \right)	\,.
\end{equation}
Although $Z_a^\chi$ diagonalizes the square $M^{\chi\dagger} M^\chi$, we need to 
perform two additional rotations in order to make $Z^{\chi T} M^\chi Z^\chi$ real and 
diagonal:
\begin{equation}
	Z_b^\chi = \left( \begin{array}{cccc} 
		1 	& 	0 	& 	0 					& 	0 \\
		0 	& 	1 	& 	0 					&	0 \\
		0 	& 	0 	&	\frac{1}{\sqrt{2}}	& 	- \frac{1}{\sqrt{2}} \\
		0 	& 	0 	& 	\frac{1}{\sqrt{2}} 	& 	\frac{1}{\sqrt{2}} 
		\end{array} \right)  \qquad \mbox{and} \qquad 
		Z_c^\chi = \left( \begin{array}{cccc} 
			e^{-i\phi_{M_1}/2} 	& 	0 					& 	0 	& 	0 \\
			0 					&	e^{-i\phi_{M_2}/2} 	& 	0 	&	0 \\
			0 					& 	0 					& 	e^{-i\phi_\mu/2} 	& 	0 \\
			0 					& 	0 					& 	0 					& 	e^{-i\phi_\mu/2} 
			\end{array} \right)\,,
\end{equation}
where $\phi_{M_i,\mu}$ is the phase of $M_{1,2}$ and $\mu$ respectively.  The 
resulting mixing matrix is given by
\begin{equation}
	Z^\chi = Z_a^\chi Z_b^\chi Z_c^\chi \,,
\end{equation}
from which we deduce the relevant components for the lightest neutralino
\begin{align}
	Z^{\chi}_{11}	&=	e^{-\tfrac{i}{2}\phi_{M_1}} + \Order(v^2/M_\mathrm{SUSY}^2) \,, \notag \\
	Z^\chi_{21} 	&=	\Order(v^2/M_\mathrm{SUSY}^2) \,, \notag \\
	Z^\chi_{31} 	&=	-\frac{e^{-\tfrac{i}{2}\phi_{M_1}}}{\sqrt{2}} \frac{g_1 v}{|M_1|^2-|\mu|^2} (M_1c_\beta + \mu^* s_\beta) 
	+ \Order(v^2/M_\mathrm{SUSY}^2)\,, \notag \\
	Z^\chi_{41}		&=	\frac{e^{-\tfrac{i}{2}\phi_{M_1}}}{\sqrt{2}} \frac{g_1 v}{|M_1|^2-|\mu|^2} (M_1s_\beta + \mu^* c_\beta) 
	+ \Order(v^2/M_\mathrm{SUSY}^2)\,.
\end{align}

\section{Analytic expressions for $\boldsymbol{\chi}$--nucleon scattering}
\label{AppB}
This Appendix concerns the derivation of the analytic expressions 
(\ref{up-squark-contribution}-\ref{eq:C Higgs dwn}) for the Wilson coefficients 
$C_{q_i}$ appearing in the scalar $\chi$--nucleon couplings (\ref{fN}).  The 
derivation involves an analysis of the spin-independent (SI) amplitude for 
$\chi q_i\to\chi q_i$ scattering due to tree-level Higgs and squark exchange 
(Fig.\ \ref{fig:spin_ind}).  

\textbf{\textit{Squark exchange---}}Consider first the contribution due to squark 
exchange, where the zero-momentum propagator for the $s$- and $u$-channels is 
denoted by
\begin{equation}
	D^\pm_{q_i\tilde{q}_s}
	= \frac{1}{(m_\chi \pm m_{q_i})^2 - m_{\tilde{q}_s}^2+i\epsilon}\,,
\end{equation}
and we define
\begin{equation}
	i( \Gamma_L^{q_i\tilde{q}_s*} P_R + \Gamma_R^{q_i\tilde{q}_s*} P_L)
\end{equation}
as the Feynman rule for the $\chi \bar{q}_i\tilde{q}_s$ coupling.  Then for 
spinors $u_i$ and $v_i$ carrying momentum $p_i$, the $s$-channel amplitude is
\begin{align}
	{\cal A}_s^{q_i} 
	&= \sum_{s=1}^6 \bar{v}_1 [i( \Gamma_L^{q_i\tilde{q}_s*} P_R 
	+ \Gamma_R^{q_i\tilde{q}_s*} P_L)] u_2
	\left[ iD^+_{q_i\tilde{q}_s} \right] \bar{u}_4 [i( \Gamma_L^{q_i\tilde{q}_s} P_L 
	+ \Gamma_R^{q_i\tilde{q}_s} P_R)] v_3  \notag\\
	&= -\frac{i}{2}\sum_{s=1}^6 D^+_{q_i\tilde{q}_s} 
	\left\{ 
	\Gamma_L^{q_i\tilde{q}_s*}\Gamma_L^{q_i\tilde{q}_s} 
	(\bar{v}_1 \gamma^\mu P_L v_3)(\bar{u}_4 \gamma_\mu P_R u_2) 
	+ \Gamma_R^{q_i\tilde{q}_s*}\Gamma_R^{q_i\tilde{q}_s} 
	(\bar{v}_1\gamma^\mu P_R v_3)(\bar{u}_4 \gamma_\mu P_L u_2) 
	 \right. \notag \\
	&\left. + \Gamma_L^{q_i\tilde{q}_s*}\Gamma_R^{q_i\tilde{q}_s} 
	\left[ (\bar{v}_1 P_R v_3)(\bar{u}_4  P_R u_2) 
	+ \tfrac{1}{4} (\bar{v}_1 \sigma^{\mu\nu} v_3)
	(\bar{u}_4 \sigma_{\mu\nu} P_R u_2) \right] \right. \notag \\
	&\left. + \Gamma_R^{q_i\tilde{q}_s*}\Gamma_L^{q_i\tilde{q}_s} 
	\left[ (\bar{v}_1 P_L v_3)(\bar{u}_4  P_L u_2) 
	+ \tfrac{1}{4} (\bar{v}_1 \sigma^{\mu\nu} v_3)
	(\bar{u}_4 \sigma_{\mu\nu} P_L u_2) \right]    \right\}\,,
\end{align}
where the Fierz identities 
\begin{align}
	(P_{L,R})_{ij}(P_{L,R})_{kl} &= \tfrac{1}{2} (P_{L,R})_{il}(P_{L,R})_{kj} 
	+ \tfrac{1}{8}(\sigma^{\mu\nu})_{il}(\sigma_{\mu\nu}P_{L,R})_{kj}\,, \notag\\
	(P_{L,R})_{ij}(P_{R,L})_{kl} &= 
	\tfrac{1}{2}(\gamma^\mu P_{R,L})_{il} (\gamma_\mu P_{L,R})_{kj}\,,
\end{align}
have been used to obtain the final equality.  Similarly, in the $u$-channel we 
find
\begin{align}
	{\cal A}_u^{q_i} 
	&= -\sum_{s=1}^6 \bar{v}_3 
	[i( \Gamma_L^{q_i\tilde{q}_s} P_L + \Gamma_R^{q_i\tilde{q}_s} P_R)] u_2 
	\left[ i D^-_{q_i\tilde{q}_s} \right] \bar{u}_4 
	[i( \Gamma_L^{q_i\tilde{q}_s*} P_R + \Gamma_R^{q_i\tilde{q}_s*} P_L)] v_1  \notag\\
	&= \frac{i}{2}\sum_{s=1}^6 D^-_{q_i\tilde{q}_s}
	\left\{ 
	\Gamma_L^{q_i\tilde{q}_s*}\Gamma_L^{q_i\tilde{q}_s} 
	(\bar{v}_3 \gamma^\mu P_R v_1)(\bar{u}_4 \gamma_\mu P_L u_2) 
	+ \Gamma_R^{q_i\tilde{q}_s*}\Gamma_R^{q_i\tilde{q}_s} 
	(\bar{v}_3 \gamma^\mu P_L v_1)(\bar{u}_4 \gamma_\mu P_R u_2) 
	 \right. \notag \\
	&\left. + \Gamma_L^{q_i\tilde{q}_s}\Gamma_R^{q_i\tilde{q}_s*} 
	\left[ (\bar{v}_3 P_L v_1)(\bar{u}_4  P_L u_2) 
	+ \tfrac{1}{4} (\bar{v}_3 \sigma^{\mu\nu} v_1)
	(\bar{u}_4 \sigma_{\mu\nu} P_L u_2) \right] \right. \notag \\
	&\left. + \Gamma_R^{q_i\tilde{q}_s}\Gamma_L^{q_i\tilde{q}_s*} 
	\left[ (\bar{v}_3 P_R v_1)(\bar{u}_4  P_R u_2) 
	+ \tfrac{1}{4} (\bar{v}_3 \sigma^{\mu\nu} v_1)
	(\bar{u}_4 \sigma_{\mu\nu} P_R u_2) \right]    \right\}\,,
\end{align}
so neglecting the spin-dependent terms involving $\gamma_\mu$ and 
$\gamma_\mu\gamma_5$ gives,
\begin{align}
	\left.{\cal A}_{s+u}^{q_i}\right|_\mathrm{SI} 
	&= \frac{i}{4} \sum_{s=1}^6 \left[ D^+_{q_i\tilde{q}_s} + D^-_{q_i\tilde{q}_s} \right] 
	\Re\left\{\Gamma_L^{q_i\tilde{q}_s}\Gamma_R^{q_i\tilde{q}_s*} \right\}  
	(\bar{v}_3  v_1)(\bar{u}_4   u_2) 
	+ \gamma_5 \mbox{ terms}\,.
	\label{eqn:spin_amp}
\end{align} 
At zero-momentum transfer, the terms involving $\gamma_5$ are suppressed and so 
we can read off the Wilson coefficient for the operator 
$\bar{\chi}\chi\bar{q}_iq_i$:
\begin{align}
	\bar{m}_{q_i} C_{q_i}^{\tilde{q}} &= \frac{1}{8} \sum_{s=1}^6 
	\left[ D^+_{q_i\tilde{q}_s} + D^-_{q_i\tilde{q}_s}\right] 
	\Re\left\{\Gamma_L^{q_i\tilde{q}_s}\Gamma_R^{q_i\tilde{q}_s*} \right\} \notag \\
	&= \frac{1}{8} \sum_{s=1}^6 \left[ 
	\frac{1}{(m_\chi + m_{q_i})^2 - m_{\tilde{q}_s}^2+i\epsilon} 
	+ \frac{1}{(m_\chi - m_{q_i})^2 - m_{\tilde{q}_s}^2+i\epsilon}\right] 
	\Re\left\{\Gamma_L^{q_i\tilde{q}_s}\Gamma_R^{q_i\tilde{q}_s*} \right\} \,,
	\label{mqCq}
\end{align}
where there is no sum over $q_i$ and $\bar{m}_{q_i}$ is the running quark mass.  
In the literature, the quark mass in $D^{\pm}_{q_i\tilde{q}_s}$ is often neglected, 
in which case the $s$- and $u$-channel amplitudes coincide with each other.  
Note that by substituting the couplings~\cite{Rosiek:1995kg}
\begin{align}
	\Gamma_L^{d_i\tilde{d}_s}	&=	\sqrt{2}g_2 (\tfrac{1}{2}Z^\chi_{21} 
	- \tfrac{1}{6}\tan\theta_W Z^\chi_{11}) Z_{is}^{\tilde{d}*} 
	- Y_{d_i}^* Z^\chi_{31}Z_{i+3,s}^{\tilde{d}*}\,, \notag\\
	\Gamma_R^{d_i\tilde{d}_s}	&=	
	-\frac{\sqrt{2}}{3}g_2\tan\theta_W Z^{\chi*}_{11}Z_{i+3,s}^{\tilde{d}*} 
	- Y_{d_i}Z^{\chi*}_{31} Z_{is}^{\tilde{d}*}\,,\notag \\
	\Gamma_L^{u_i\tilde{u}_s}	&= 	-\sqrt{2}g_2 (\tfrac{1}{2}Z^\chi_{21} 
	+\tfrac{1}{6}\tan\theta_W Z^\chi_{11}) Z_{is}^{\tilde{u}*} 
	- Y_{u_i}^* Z^\chi_{41}Z_{i+3,s}^{\tilde{u}*} \,,\notag \\
	\Gamma_R^{u_i\tilde{u}_s}	&= 
	\frac{2\sqrt{2}}{3}g_2\tan\theta_W Z^{\chi*}_{11} Z_{i+3,s}^{\tilde{u}*} 
	- Y_{u_i} Z^{\chi*}_{41} Z_{is}^{\tilde{u}*}\,,
\end{align}
into (\ref{mqCq}), one recovers the expressions given in~\cite{Falk:1998xj}.

To simplify (\ref{mqCq}), we expand all mixing matrices in powers of 
$v/M_\mathrm{SUSY}$.  At leading order, the elements $Z_{IJ}^\chi$ are given by 
(\ref{eq:Zchi}), while products of squark mixing matrices simplify as 
follows~\cite{Crivellin:2012zz}:
\begin{align}
	\sum_{s=1}^6 Z_{is}^{\tilde{u}}Z_{i+3,s}^{\tilde{u}*}
	D^{\pm}_{q_i\tilde{q}_s}
	&= \Delta_{u_i} 	\frac{L_{u_i}^{\pm} - R_{u_i}^{\pm}}
	{m_{\tilde{u}_i^L}^2 - m_{\tilde{u}_i^R}^2} \notag\\
	\sum_{s=1}^6 Z_{is}^{\tilde{d}}Z_{i+3,s}^{\tilde{d}*} 
	D_{q_i\tilde{q}_s}^{\pm}
	&= \Delta_{d_i} \frac{ L_{d_i}^{\pm} - R_{d_i}^{\pm}}
	{m_{\tilde{d}_i^L}^2 - m_{\tilde{d}_i^R}^2} \notag\\
	\sum_{s=1}^6 Z_{i+3,s}^{\tilde{d}}Z_{i+3,s}^{\tilde{d}*} 
	D_{q_i\tilde{q}_s}^{\pm} 
	&= R_{d_i}^{\pm} \notag\\
	\sum_{s=1}^6 Z_{is}^{\tilde{d}}Z_{is}^{\tilde{d}*} 
	D_{q_i\tilde{q}_s}^{\pm}  
	&= L_{d_i}^{\pm} \notag\\
	\sum_{s=1}^6 Z_{i+3,s}^{\tilde{u}}Z_{i+3,s}^{\tilde{u}*} 
	D_{q_i\tilde{q}_s}^{\pm}
	&= R_{u_i}^{\pm} \notag\\
	\sum_{s=1}^6 Z_{is}^{\tilde{u}}Z_{is}^{\tilde{u}*} 
	D_{q_i\tilde{q}_s}^{\pm}
	&= L_{u_i}^{\pm}\,,
\end{align}
where $m_{\tilde{q}_i^{L}}^2$ and $m_{\tilde{q}_i^R}^2$ correspond to the upper 
and lower diagonal elements of the squark (mass)$^2$ matrices (\ref{eq:sqrk mass}), 
\begin{align}
	\Delta_{u_i} = -\bar{m}_{u_i}(A^{ii}_u + \mu\cot\beta) \qquad 
	\mbox{and} \qquad \Delta_{d_i} = -\bar{m}_{d_i}(A^{ii}_d + \mu \tan\beta)\,,
\end{align}
are the off-diagonal elements, and the squark propagators in the chiral basis are
\begin{equation}
	S_{q_i}^{\pm} = \frac{1}{(m_\chi\pm m_{q_i})^2 - m_{\tilde{q}_i^S}^2+i\epsilon} \,, 
	\qquad \mbox{for } S=L \mbox{ or } R\,.
\end{equation}
Neglecting terms of $\Order(v/M_\mathrm{SUSY}^2)$, the final result is
\begin{align}
	\bar{m}_{u_i} C_{u_i}^{\tilde{q}} &= \frac{1}{8}\Re \bigg\{ 
	-\frac{4}{3}g_2^2 \tan\theta_W(\tfrac{1}{2}Z^{\chi*}_{11} Z^{\chi*}_{21} 
	+ \tfrac{1}{6}\tan\theta_W Z^{\chi*}_{11} Z^{\chi*}_{11}) \Delta_{u_i} 
	L_{u_i}^+ R_{u_i}^+  \notag \\
	&-\frac{2\sqrt{2}}{3} g_2 \tan\theta_W Z^{\chi*}_{11} Z^{\chi*}_{41} 
	Y_{u_i}R_{u_i}^+ 
	+ \sqrt{2}g_2(\tfrac{1}{2} Z^{\chi*}_{21} Z^{\chi*}_{41} 
	+ \tfrac{1}{6} \tan\theta_W Z^{\chi*}_{11} Z^{\chi*}_{41}) 
	Y_{u_i} L_{u_i}^+\bigg\} 
	\notag \\
	&+ (L^+,R^+) \leftrightarrow (L^-,R^-) \notag\\
	&= \bar{m}_{u_i}\frac{g_1^2}{8}  \Re \left\{ e^{i\phi_{M_1}} \left[ 
	\frac{2}{9} X_{u_i} L_{u_i}^+ R_{u_i}^+ 
	+ \frac{1}{6} \frac{(M_1^* + \mu\cot\beta)}{|M_1|^2-|\mu|^2} 
	(L_{u_i}^+ - 4R_{u_i}^+ ) \right]\right\} \notag \\
	&+ (L^+,R^+) \leftrightarrow (L^-,R^-) \,, \notag\\
	\bar{m}_{d_i}C_{d_i}^{\tilde{q}} &= \frac{1}{8}\Re \left\{ 
	-\frac{2}{3}g_2^2\tan\theta_W (\tfrac{1}{2} Z^{\chi*}_{21} Z^{\chi*}_{11} 
	- \tfrac{1}{6}\tan\theta_W Z^{\chi*}_{11} Z^{\chi*}_{11}) 
	\Delta_{d_i} L_{d_i}^+ R_{d_i}^+	\right.\notag \\
	&\left. -\sqrt{2}g_2(\tfrac{1}{2} Z^{\chi*}_{21} Z^{\chi*}_{31} 
	- \tfrac{1}{6}\tan\theta_W Z^{\chi*}_{11} Z^{\chi*}_{31}) Y_{d_i} L_{d_i}^+ 	
	+\frac{\sqrt{2}}{3}g_2 \tan\theta_W Z^{\chi*}_{11} Z^{\chi*}_{31} 
	Y_{d_i} R_{d_i}^+\right\} \notag \\
	&+ (L^+,R^+) \leftrightarrow (L^-,R^-)\notag \\
	&= -\bar{m}_{d_i}\frac{g_1^2}{8} \Re \left\{ e^{i\phi_{M_1}} \left[ 
	\frac{1}{9} X_{d_i} L_{d_i}^+ R_{d_i}^+
	+ \frac{1}{6} \frac{(M_1^* + \mu\tan\beta)}{|M_1|^2-|\mu|^2} 
	(L_{d_i}^+ + 2  R_{d_i}^+) \right] \right\} \notag \\
	&+ (L^+,R^+) \leftrightarrow (L^-,R^-) \,,
\end{align}
where the squark mixing $X_{q_i}$ is defined in (\ref{Xq}).

\textbf{\textit{Higgs exchange---}}Now consider the Higgs contribution, for 
which the $t$-channel amplitude reads
\begin{align}
	{\cal A}_t^{h,H} 	&= 	\sum_{k=1,2} \bar{u}_3 \left[ 
	i(\Gamma_{\chi\chi}^{H_k}P_R + \Gamma_{\chi\chi}^{H_k*}P_L) \right] u_1 
	\left[\frac{i}{-m_{H_k}^2}\right] \bar{u}_4 \left[ i(\Gamma_{q_iq_i}^{H_k}P_R 
	+ \Gamma_{q_iq_i}^{H_k*}P_L) \right] u_2  \notag\\
	&= i \sum_{k=1,2}\frac{1}{m_{H_k}^2} \Re\{\Gamma_{\chi\chi}^{H_k}\}\Re\{
	\Gamma_{q_iq_i}^{H_k}\} (\bar{u}_3u_1)(\bar{u}_4u_2) 
	+ \mbox{spin-dependent terms}\,.
\end{align}
Evidently, the Wilson coefficient due to Higgs exchange is
\begin{equation}
	C_{q_i}^{h,H} = \frac{1}{2}\sum_{k=1}^2\frac{1}{m_{H_k}^2}
	\Re\{\Gamma_{\chi\chi}^{H_k}\}\Re\{\Gamma_{q_iq_i}^{H_k}\}\,,
\end{equation}
and as in the case for squark exchange, we substitute the $H_k^0\chi\chi$ and 
$H_k^0q_iq_i$ couplings~\cite{Rosiek:1995kg}
\begin{align}
	\Gamma_{\chi\chi}^{H_k} 	&=	\frac{g_2}{c_W} (Z^h_{1k} Z^{\chi*}_{31} - Z^h_{2k} Z^{\chi*}_{41}) (Z^{\chi}_{11}s_W - Z^{\chi}_{21}c_W) \,, \notag\\
	\Gamma_{q_iq_i}^{H_k} 		&=	-\frac{Y_{q_i}}{\sqrt{2}}Z^h_{qk} \qquad 
	\mbox{where } Z^h_{qk} = 
	\left\{\begin{array}{cl} 
	Z^h_{2k} & \mbox{for }q=u \\
	-Z^h_{1k} & \mbox{for }q=d 
	\end{array}\right. \,,
\end{align}
to obtain the analytic expressions
\begin{align}
	\bar{m}_{u_i}C_{u_i}^{h,H} &= \frac{g_2}{2\sqrt{2}}\Re\{Y_{u_i}\} 
	\left[ \Re\{Z^{\chi*}_{41}	(Z^\chi_{11}\tan\theta_W - Z^\chi_{21})\} 
	\left(\frac{c_\alpha^2}{m_h^2}+\frac{s_\alpha^2}{m_H^2} \right) \right. 
	\notag \\
	&\left.+  \Re\{Z^{\chi*}_{31}(Z^\chi_{11}\tan\theta_W - Z^\chi_{21})\} 
	s_\alpha c_\alpha\left(\frac{1}{m_h^2} - \frac{1}{m_H^2} \right) \right]\notag \\
	&= \frac{g_1^2}{4}\frac{\bar{m}_{u_i}}{|M_1|^2-|\mu|^2} \left[ 
	\Re\{M_1^* + \mu\cot\beta\} \left(\frac{c_\alpha^2}{m_h^2}
	+\frac{s_\alpha^2}{m_H^2} \right) \right.\notag \\
	&\left.- \Re\{M_1^*\cot\beta + \mu\}s_\alpha c_\alpha \left(\frac{1}{m_h^2} 
	- \frac{1}{m_H^2} \right) \right] \notag\\
	\bar{m}_{d_i}C_{d_i}^{h,H} &= \frac{g_2}{2\sqrt{2}}
	\Re\{Y_{d_i}\} \left[ \Re\{Z^{\chi*}_{31} 
	(Z^\chi_{11}\tan\theta_W - Z^\chi_{21})\} \left(\frac{s_\alpha^2}{m_h^2} 
	+\frac{c_\alpha^2}{m_H^2} \right) \right. \notag \\
	&\left.+  \Re\{Z^{\chi*}_{41} (Z^\chi_{11} \tan\theta_W - Z^\chi_{21})\} 
	s_\alpha c_\alpha\left(\frac{1}{m_h^2} - \frac{1}{m_H^2} \right) \right] \notag\\
	&= \frac{g_1^2}{4}\frac{\bar{m}_{d_i}}{|M_1|^2-|\mu|^2} \left[ 
	\Re\{M_1^* + \mu\tan\beta\} \left(\frac{s_\alpha^2}{m_h^2}
	+\frac{c_\alpha^2}{m_H^2} \right) \right.\notag \\
	&\left.- \Re\{M_1^*\tan\beta + \mu\} s_\alpha c_\alpha \left(\frac{1}{m_h^2} 
	- \frac{1}{m_H^2} \right) \right]\,.
\end{align}
 
\bibliography{BIB_CHPT}
\end{document}